\documentclass[twocolumn,twocolappendix]{aastex631}

\received{\today}
\submitjournal{ApJ}

\shorttitle{The AGN LF at $z\sim 4$}
\shortauthors{Boutsia et al.}
\graphicspath{{./}{figures/}}

\begin{document}

\title{A nearly complete census of ultra bright QSOs at $z\sim 4$ in the southern hemisphere: the AGN Luminosity Function}

\correspondingauthor{Boutsia Konstantina}
\email{konstantina.boutsia@noirlab.edu}

\author[0000-0003-4432-5037]{Konstantina Boutsia}
\affil{Cerro Tololo Inter-American Observatory/NSF NOIRLab, Casilla 603, La Serena, Chile}

\author[0000-0002-5688-0663]{Andrea Grazian}
\affil{INAF--Osservatorio Astronomico di Padova, 
Vicolo dell'Osservatorio 5, I-35122, Padova, Italy}

\author[0000-0003-0734-1273]{Emanuele Giallongo}
\affil{INAF--Osservatorio Astronomico di Roma, Via Frascati 33, I-00078,
Monte Porzio Catone, Italy}

\author[0000-0002-2115-5234]{Stefano Cristiani}
\affil{INAF--Osservatorio Astronomico di Trieste, 
Via G.B. Tiepolo, 11, I-34143, Trieste, Italy}
\affiliation{INFN-National Institute for Nuclear Physics,  
via Valerio 2, I-34127 Trieste}
\affil{IFPU--Institute for Fundamental Physics of the Universe,
via Beirut 2, I-34151, Trieste, Italy}

\author[0000-0003-4744-0188]{Fabio Fontanot}
\affil{INAF--Osservatorio Astronomico di Trieste, 
Via G.B. Tiepolo, 11, I-34143, Trieste, Italy} 
\affil{IFPU--Institute for Fundamental Physics of the Universe,
via Beirut 2, I-34151, Trieste, Italy}

\author[0000-0002-7738-5389]{Giorgio Calderone}
\affil{INAF--Osservatorio Astronomico di Trieste, 
Via G.B. Tiepolo, 11, I-34143, Trieste, Italy}

\author[0000-0003-3693-3091]{Valentina D'Odorico}
\affil{INAF--Osservatorio Astronomico di Trieste, 
Via G.B. Tiepolo, 11, I-34143, Trieste, Italy} 
\affil{IFPU--Institute for Fundamental Physics of the Universe,
via Beirut 2, I-34151, Trieste, Italy}

\author[0000-0003-4740-9762]{Francesco Guarneri}
\affil{Hamburger Sternwarte, Universitat Hamburg, Gojenbergsweg 112, D-21029 Hamburg, Germany}

\author[0009-0004-2597-6146]{Matteo Porru}
\affil{INAF--Osservatorio Astronomico di Trieste, 
Via G.B. Tiepolo, 11, I-34143, Trieste, Italy} 
\affil{Dipartimento di Fisica, Sezione di Astronomia, Universit\`a di Trieste,
via G.B. Tiepolo 11, I-34131, Trieste, Italy}

\author[0000-0002-5959-5964]{Andrea Trost}
\affil{INAF--Osservatorio Astronomico di Trieste, 
Via G.B. Tiepolo, 11, I-34143, Trieste, Italy} 
\affil{Dipartimento di Fisica, Sezione di Astronomia, Universit\`a di Trieste,
via G.B. Tiepolo 11, I-34131, Trieste, Italy}
\affil{SISSA --  International School for Advanced Studies, Via Bonomea 265, I-34136 Trieste, Italy}
\affil{INFN -- National Institute for Nuclear Physics, via Valerio 2, I-34127 Trieste}

\begin{abstract}
We present a significantly extended and nearly complete determination of the Active Galactic Nuclei (AGN) Luminosity Function (LF) at redshift $z\sim4$, spanning more than 10 magnitudes in absolute luminosities ($M_{1450}$ = -30.0 to -18.0). 
We combine a spectroscopically complete, bright-QSO sample from QUBRICS for Y-band$\leq$18.5 ($\sim M_{1450} \leq -27.5$) in the 3.6$\leq z \leq$4.2 redshift range, over 13,966 $deg^{2}$, with intermediate-luminosity data from DESI DR1 and faint-end constraints from JWST. 
This allows us to provide the most stringent constraints to date on the `knee'  ($M_{1450}^{*}$) of the LF offering a highly accurate description of the transition between bright and faint AGN populations. Including recent JWST measurements yields higher faint-AGN space densities than most pre-JWST determinations and a fitted faint-end slope of $\alpha=-1.92$. This regime, which has important implications for understanding early supermassive black hole (SMBH)
growth and their impact on cosmological Reionization, nevertheless remains subject to substantial field-to-field variance and AGN-classification uncertainties.
Comparison with the GAEA semi-analytic model indicates that while theoretical predictions align with faint-end observations, they tend to overpredict the speed of redshift evolution at the bright end between $z\sim$4 and $z\sim$5.
Our estimate for the $z\sim$4 LF represents a critical benchmark for studying supermassive black hole (SMBH) evolution and making accurate predictions for future wide-area imaging surveys such as Euclid, Vera Rubin LSST, and Roman.
\end{abstract}

\keywords{Observational cosmology (1146); Quasars (1319); Catalogs (205); Surveys (1671)}

\section{Introduction}
The study of the AGN population, and in particular of its brightest counterparts -  i.e. quasars (QSOs)  at very early cosmic epochs 
is crucial to understanding supermassive black hole (SMBH) formation and evolution, the intergalactic medium (IGM), and cosmological reionization, \citep[e.g.][]{volonteri23, Dodorico23, fontanot23, Madau24}.
One of the most important observational indicators for the evolution of the AGN/QSO population is the QSO luminosity function (LF) which has been the subject of intense study and revision in recent years \citep[e.g.][]{fan22}. In the high-$z$ regime, the shape of the QSO LF, usually parametrized as a double power-law function, is a highly debated topic, and recent observations from both wide ground-based and deep space-based surveys have significantly advanced our understanding. Despite current efforts to constrain the QSO LF at high-$z$ - e.g. at $z\sim$6 \citep{Jiang2016, Matsuoka23, Yang2019} - and even close to $z\sim$8 \citep{Mori20}, there is still no consensus on the detailed slope of the LF at 3$<z<$5, both at the bright and faint-end \citep{Akiyama18,shen20,boutsia21}. Therefore, a detailed census
of AGNs, both at bright (M$_{1450}<$-26) and faint magnitudes (M$_{1450}\geq$-22) at $z\sim$4, can give interesting constraints on the evolution of the population.

At bright magnitudes, the Sloan Digital Sky Survey (SDSS) constituted an unprecedented milestone for the space density of bright QSOs at $z>3$ for at least 20 years \citep[e.g. SDSS DR18][]{Almeida23}. However, recent surveys have established that the number of bright QSOs at $z>$3 was previously underestimated by 30\% to 40\% \citep{boutsia18,Sch19a}. Such efforts, like the Extremely Luminous QSO Survey (ELQS) and QUBRICS (QUasars as BRIght beacons for Cosmology in the Southern hemisphere), are addressing completeness issues and exploring previously unstudied regions of the sky to better constrain the QSO LF at various redshifts. The QUBRICS survey, specifically, focuses on the Southern Hemisphere to identify very bright QSOs at $z>2.5$ aiming to improve our understanding of the QSO LF in regions and magnitudes that have been underestimated or unexplored by previous studies. 

On the faint-end of the AGN LF, JWST has revolutionized our knowledge by discovering numerous low-luminosity AGNs that were inaccessible to ground-based telescopes \citep{kocevski23,Harikane23,maiolino23lf, Taylor25, Juod26}. The number density of these faint AGNs is substantially higher than extrapolations based on previous QSO LFs \citep{shen20}, reaching $~10^{-5}$ cMpc$^{-3}$. This confirms earlier hints based on deep HST and Chandra surveys \citep{Giallongo15,giallongo19}. A notable subset of these faint AGNs are `Little Red Dots' (LRDs), characterized by compact morphologies and peculiar `V-shaped' spectral energy distributions (SEDs) with red optical and blue UV slopes \citep{LRD}. LRDs may represent an early, dust-obscured phase of black hole growth \citep[e.g.][]{kocevski24}. Because the dominant power source of LRDs (AGN versus stellar) is still debated, we do not re-classify LRDs in this work. In our LF compilation we simply adopt the AGN selections and space densities published by the cited JWST studies that still present considerable scatter for faint AGNs. 
This has led to different models for the faint end of the AGN LF at $z\sim 5$, described as `conservative', `intermediate', and `high' options, depending on which data sets are included \citep{Grazian24}.

Combining the new data from both bright QSOs and faint AGNs provides a more complete picture of the AGN LF at $z\sim$4.
When combining bright data from QUBRICS with fainter data from the literature, there is evidence for a differential evolution, where the bright end evolves more rapidly with redshift than the faint end, which appears to remain almost constant from $z$=4 to $z$=10 \citep{Grazian24}. State-of-the-art semi-analytic models like GAEA \citep{DeLucia24}, DELPHI \citep{dayal24} and CAT \citep{trinca22} 
can reasonably reproduce the observed AGN LF at $z>4$, particularly at the faint end. However, the models tend to predict a faster redshift evolution than the one observed, especially for intermediate-luminosity AGNs. 

Despite these recent advances, significant uncertainties remain about the shape of the AGN LF at $z\ge 4$ and its evolution.
In order to systematize the knowledge of the bright side of the QSO LF at $z\sim 4$, in this work we have significantly extended the collection of bright QSOs in the 3.6$<z<$4.2 redshift range, in order to compile the most complete LF that spans a wide range of absolute magnitudes in the literature. 
Using data from the recent Dark Energy Spectroscopic Instrument (DESI) DR1 release (S. Juneau, et al. (in prep.)), we are able to extend the $z\sim 4$ QSO LF by approximately two magnitudes, i.e. down to M$_{1450}$= -25, with respect to the one of \citet{boutsia21}.
Thanks to recent JWST surveys of AGNs at $z\sim 4$ \citep[e.g.][]{maiolino23lf}, we can further extend the AGN LF down to an absolute magnitude of M$_{1450}$= -18.

The structure of this paper is the following: in Section \ref{sec:data}, we summarize the photometric and spectroscopic data of QSOs at $z\sim 4$ used in this work. In Section \ref{sec:method}, we describe the method adopted to derive the AGN LF, and in Section \ref{sec:bestFit}, we show the results we obtained. Comparison of the AGN LF at $z\sim 4$ with the recent results at higher redshifts and with model predictions is shown in Section \ref{sec:comparison}. The discussion and conclusions are summarized in Sections \ref{sec:discussion} and \ref{sec:conclusions}. Unless otherwise stated, apparent magnitudes are in the AB photometric system. We adopt here the $\Lambda$-CDM concordance cosmological model, with H$_{0}$=70 km s$^{-1}$ Mpc$^{-1}$, $\Omega_{M}$=0.3, and $\Omega_{\Lambda}$=0.7.


\section{Data}
\label{sec:data}

In order to calculate the LF of QSOs at $z\sim$4, we used the data from two surveys, QUBRICS and DESI.
 
The QUBRICS collaboration has been active since 2019, selecting and spectroscopically confirming bright ($i<$18.5) and high-redshift ($z>$2.5) QSOs in the Southern Hemisphere \citep{calderone19,boutsia20,boutsia21,Guarneri21,guarneri22,grazian22,cupani22,Cristiani23,calderone23,Porru26}. 
The selection algorithm, based on machine learning techniques, has been evolving along with the spectroscopic follow-up campaigns, using all newly confirmed QSOs as an improved training set to select candidates with greater precision. In addition to constantly updating the training set, various machine learning methods have also been explored. After starting with a Canonical Correlation Analysis \citep[CCA,][]{CCA}, we moved to a Probabilistic Random Forest \citep[PRF,][]{PRF2019} approach that builds on the Random Forest algorithm and is better suited to dealing with incomplete data-sets and properly accounting for measurement errors (e.g. photometric upper limits). In addition, the CCA was based on magnitudes for the classification, while the PRF and all subsequent methods use color information. This feature makes the selection methods more efficient and minimizes the bias against bright objects. 

The most recent version of our selection algorithm is based on a technique called gradient boosting  \citep[XGBoost,][]{XGBoost}. This algorithm has been modified into a new technique called `Boost recall' or `inverse selection' which narrows down the candidate pool by selecting and rejecting non-interesting sources in an iterative way \citep{calderone23}. At the end of this selection process, the remaining dataset should include only bright, high-$z$ QSOs. The boost-recall algorithm increases completeness while maintaining a high success rate and in addition produces a redshift estimate that is very useful in prioritizing candidate follow-up observations. Details on the evolution of the adopted selection algorithms can be found in \cite{guarneri22,calderone23}.  

\subsection{The QUBRICS Main Sample}
All candidates are part of the QUBRICS Main Sample, which includes all sources for which we have photometric information in the various catalogs. Specifically, to create the current sample, we have used PanSTARRS1 \citep[DR2,][]{PS1,PSDR2} and DES \citep[Y3 Gold,][]{des21,desY3} catalogs. All sources were then cross-matched with the Gaia DR3 \citep{DeAngeli} and ALLWISE \citep{Wri10} catalogs, with a matching radius of 0.5" in both cases. The Main Sample includes millions of sources, of which $\sim$90\% are stars based on Gaia parallax and proper motion information. For the remaining 10\%, public catalogs have been used to assign spectroscopic information (e.g. Veron, SDSS, 2QZ). When this information was not available, we carried out spectroscopic follow-up in a number of facilities 
including Magellan, NTT, TNG, VLT, and LBTO. Details on the instruments used and the specific setups can be found in \cite{boutsia21}. So far, the QUBRICS Survey has confirmed $>$2000 bright, high-z QSOs in the Southern Hemisphere, finding the brightest cosmological objects in the Universe \citep{Cristiani23}. 

In this work, we collected all confirmed QSOs in our Main Sample that have a spectroscopic redshift between 3.6 and 4.2 and a magnitude limit of $Y-band\le 20.5$ in the DES and PanSTARRS catalogs. As a first step, for a consistency check, the two samples were treated separately and two luminosity functions were calculated. For the DES LF we have used all sources, and for the PanSTARRS subsample we selected only sources that have declinations $\delta<+14$ ${\deg}$, since the spectroscopic follow-up of our candidates has been limited to the Southern Hemisphere. 
The area for each survey has been calculated by counting how many HEALPix\footnote{\url{https://healpix.sourceforge.io/}} pixels contain at least one of our sources and multiplying this number by the area of a single pixel. We ran this analysis at different resolutions with HEALPix order ranging from 0 to 13, and picked the area showing the smallest relative difference between an order and the preceding one.  The resulting area estimates are 5,083 deg$^{2}$ for DES and 11,000 deg$^{2}$ for PanSTARRS. However, the two areas are partially overlapping and this was taken into consideration when calculating the `global' LF, which was created by merging the DES and PanSTARRS catalogs.

Sources in common have been included in the global catalog, considering the redshift and $Y-band$ magnitudes provided in the PanSTARRS catalog, since it is the one with the larger number of sources. We checked that for the sources in common, the DES $Y-band$ photometry was consistent with the one from PanSTARRS. The area for this merged sample was calculated again in order to account for the overlapping regions. The total number of sources used for the global LF is 2,220 (see Table \ref{tab:glob} for the ones used at the final fit) covering a total area of 13,966 deg$^{2}$.

\begin{figure}
\includegraphics[width=\linewidth]{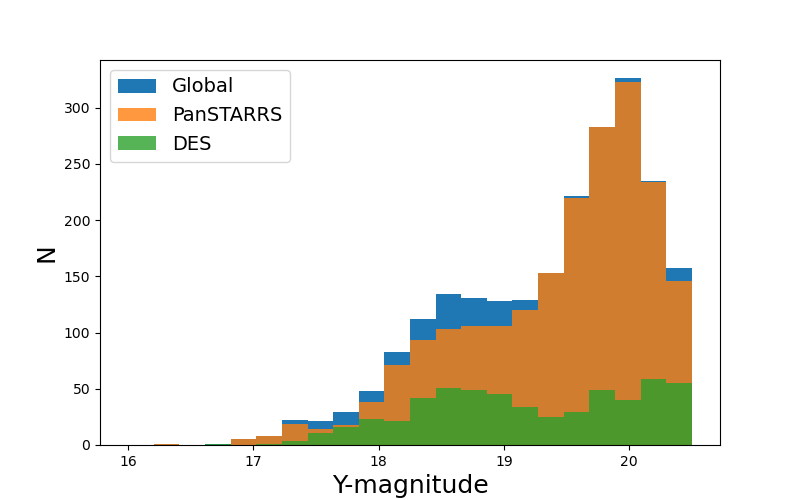}
\caption{Histogram of the Y-band magnitude of the confirmed QSOs in the PanSTARRS (orange), DES (green), and Global (blue) datasets. Notice the plateau at Y-band$>$18.5. This is an indication that our sample is not complete at magnitudes fainter than this.}
\label{fig:Hglobal}
\end{figure}

The spectroscopic confirmation of QSO candidates in the DES and PanSTARRS regions, by our group, was focused mainly on objects brighter than $Y-band$=18.5, where  
very few QSO candidates at $3.6<z<4.2$ still lack spectroscopic confirmation, based on our analysis.
Thus, we can consider that our spectroscopic follow-up is complete down to a magnitude of $Y-band$=18.5 and this can be seen in the histogram of QUBRICS sources (see Fig.\ref{fig:Hglobal}), that flattens at fainter magnitudes. Since the sources with fainter magnitudes come from both our spectroscopic follow-up, which is still ongoing, and from the literature, we find a secondary peak at $Y-band$ = 20.0. 
Since our spectroscopic sample is based on a number of publicly available data and our follow-up and selection algorithms are constantly evolving, it is very difficult to robustly estimate a selection function. For this reason, we decided to only consider QSOs with $Y-band\le$18.5, roughly corresponding to M$_{1450}$=-27.5 at $z\sim 4$, for calculating the bright-end slope of the AGN LF. 
This choice ensures that the bright-end slope is derived in the regime where spectroscopic completeness is highest and selection completeness has been independently quantified.
In fact, the completeness of the QUBRICS candidate sample has recently been calculated by \cite{Porru26} and was found to be of the order of 82\% (more details in Section 6.2). 

\subsection{DESI}

In an effort to extend the AGN LF to fainter magnitudes, we included publicly available spectra from the DESI survey \citep[DESI][]{DESIearly}. In order to have good control over the area, it was decided to focus on only five pointings that have good coverage, i.e. containing a large number of spectroscopic identifications. In this area, we selected all sources with a spectroscopic redshift in the range 3.6$<z<$4.2 in the DR1 catalog that 
have a robust spectroscopic classification (flag ZWARN=0). This resulted in 110 selected sources, of which one has a spectroscopic classification (SPECTYPE) as Galaxy and all the rest are QSOs. Then, all sources were cross-matched with the photometric catalog of PanSTARRS in order to obtain the $Y-band$ magnitude. Some sources are fainter than the PanSTARRS magnitude limit, so they have been discarded. After the crossmatch with the available photometry, the list of sources with DESI spectroscopy contains 54 sources. We also checked the source with the `Galaxy' classification (DESI ID: 39627776298324745). It is clear from the spectrum appearing in the DESI database, that this sources is a QSO, missclassified as a Galaxy. In addition, an earlier SDSS spectrum confirms it as a QSO at z=3.715 \citep{SDSS2009}.  For this reason we keep it in the list of sources to be used for the LF. 

\begin{table}
\caption{DESI fields used in our computation. We report the center coordinates and the number of considered sources. The coordinates refer to the center of the rosette, following the EDR convention \citep{EDR2024}}
\label{tab:DESIfields}
\begin{center}
\begin{tabular}{c c c c}
\hline
\hline
Rosette & RA [deg] & Dec [deg] & Nr. of sources \\
\hline
SV3R5 & 210.00 & 5.000 &  11 \\
SV3R10 &218.05 & 2.430 &  15 \\
SV3R8 & 216.30 & -0.600 &  12 \\
SV3R9 & 219.80 & -0.600 &  13 \\
SV3R17 & 212.80& -0.600 &  6 \\
\hline
\end{tabular}
\end{center}
\end{table}

For each pointing of the DESI footprint the areal coverage is 7.44 $deg.^{2}$, after accounting for the filling factor of 92.0\%, according to \cite{DESI0717}. Thus, the total effective area used for determining the LF is 37.2 deg$^{2}$, in five independent pointings the central coordinates of which can be seen in Table \ref{tab:DESIfields}, along with the number of sources per field that were considered for the calculation of the LF.

\begin{figure}
\includegraphics[width=\linewidth]{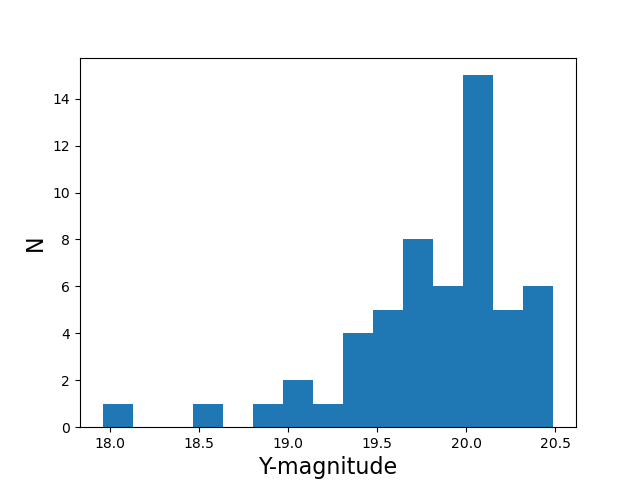}
\caption{Histogram of the Y-band magnitude of confirmed QSOs in the DESI sample adopted in this work.}
\label{fig:Hdesi}
\end{figure}

In order to check the completeness of the DESI spectroscopic campaign, we also compared it with the confirmed sources from the QUBRICS survey in the same area (the five pointings described above). 
We find two sources included in our Main Sample from the PanSTARRS survey that are not in the DESI catalogs. Thus, the final catalog that we use for calculating the LF in the DESI area includes 57 sources (see Table \ref{tab:desi}). Considering the two sources that are not included in the DESI catalog and the source that has been erroneously classified as `Galaxy', the completeness of the DESI sample is of the order of 95\% (3 sources missing out of 57). However, since we included these 3 sources in the calculation of the LF for the DESI sample, we will not correct it for completeness.

The DESI sample is more scarce compared to the PanSTARRS and DES samples, since we concentrated in a very limited area, resulting in a low number of sources (see Fig.\ref{fig:Hdesi}). 
However, we do not observe a clear plateau in the magnitude distribution. 
Since the QUBRICS sources were included in the DESI sample used for the LF calculation, we will not apply any completeness correction to the DESI sample up to an absolute magnitude of M$_{1450}$=-25.5 (corresponding to 18.0$\leq Y-band\leq$ 20.2) considering these space densities complete.


\section{Method}
\label{sec:method}

For each source, we estimate the absolute magnitude at 1450 {\AA} rest frame ($M_{1450}$) from the apparent $Y-band$ magnitude (taken from PanSTARRS when available - i.e. for 93\% of the sample - or from DES otherwise) and the spectroscopic redshift as:

\begin{equation}
M_{1450} = Y - 5 log (d_L) + 5 + 2.5 log (1 + z_{spec}) + K_{corr} \,
\end{equation}
\noindent
where $d_L$ is the luminosity distance in parsec (pc) and the color term of the k-correction $K_{corr}$ is given by the expression:
\begin{equation}
K_{corr} = -2.5\alpha_\nu log((\lambda_{obs}/(1 + z_{spec})/\lambda_{rest}) \,
\end{equation}
\noindent
where we assume a typical spectral slope for QSOs corresponding to $\alpha_\nu = 0.7$ (i.e. $F_\nu\propto \nu^{-\alpha_\nu}$), $\lambda_{rest} = 1450$ {\AA}, and $\lambda_{obs} = 9622$ {\AA} is the central wavelength of the Y-band filter. 
The adopted spectral slope, $\alpha_\nu$ = 0.7, is consistent with the median slope of optically selected QSOs at these redshifts. \citet{Rakshit20}, using the SDSS DR14Q catalog of more than 600000 quasars in the redshift range 3.6 $< z < $4.2, report a scaled median absolute deviation of MAD = 0.45 in $\alpha_\nu$, corresponding to a scatter of only DM$_{1450} \le$ 0.15 mag, well within our 0.5-mag bin width. This scatter does not significantly affect the shape of the LF.

\subsection{Number Density Determination}

The space density of bright QSOs at $z\sim$ 4 has been derived by adopting the 1/Vmax approach \citep{Eales1993}, where the
accessible volume for each object has been calculated from the
redshift interval that encompasses 3.6 $\leq z \leq$ 4.2 and the magnitude limits of the survey. 
The total area of the QUBRICS survey adopted here is 13,966 deg$^{2}$, and the sample is limited to 14.0 $\leq Y-band \leq$ 20.5. 
Only robustly confirmed QSOs \citep[i.e., flag=A, see][]{boutsia21} in the redshift interval 3.6 $\leq z \leq$ 4.2 have been selected and were used in the LF calculation (see Table \ref{tab:glob}). 
The error  bars for the QSO space density have been computed by adopting the statistics of \cite{Gehrels86}, 
which are particularly suited for low number counts, while they are close to Poisson for large numbers.

We compute the bright end of the QSO space density in ten independent intervals of absolute magnitude
from the brightest limit M$_{1450}$= -30.0 to M$_{1450}$= -25, in bins of 0.5 magnitudes. We set the faintest absolute magnitude limit to $M_{1450} = -25$, 
which is the luminosity of an object at $Y-band$ = 20.5 at $z$ = 4.2, the redshift limit of our survey. As we already discussed in the previous sections, the QUBRICS survey is not complete at these magnitudes, but we decided to extend the calculations to these faint limits in order to compare the different completeness levels of the QUBRICS and DESI samples. 

\begin{figure}
\includegraphics[width=\linewidth]{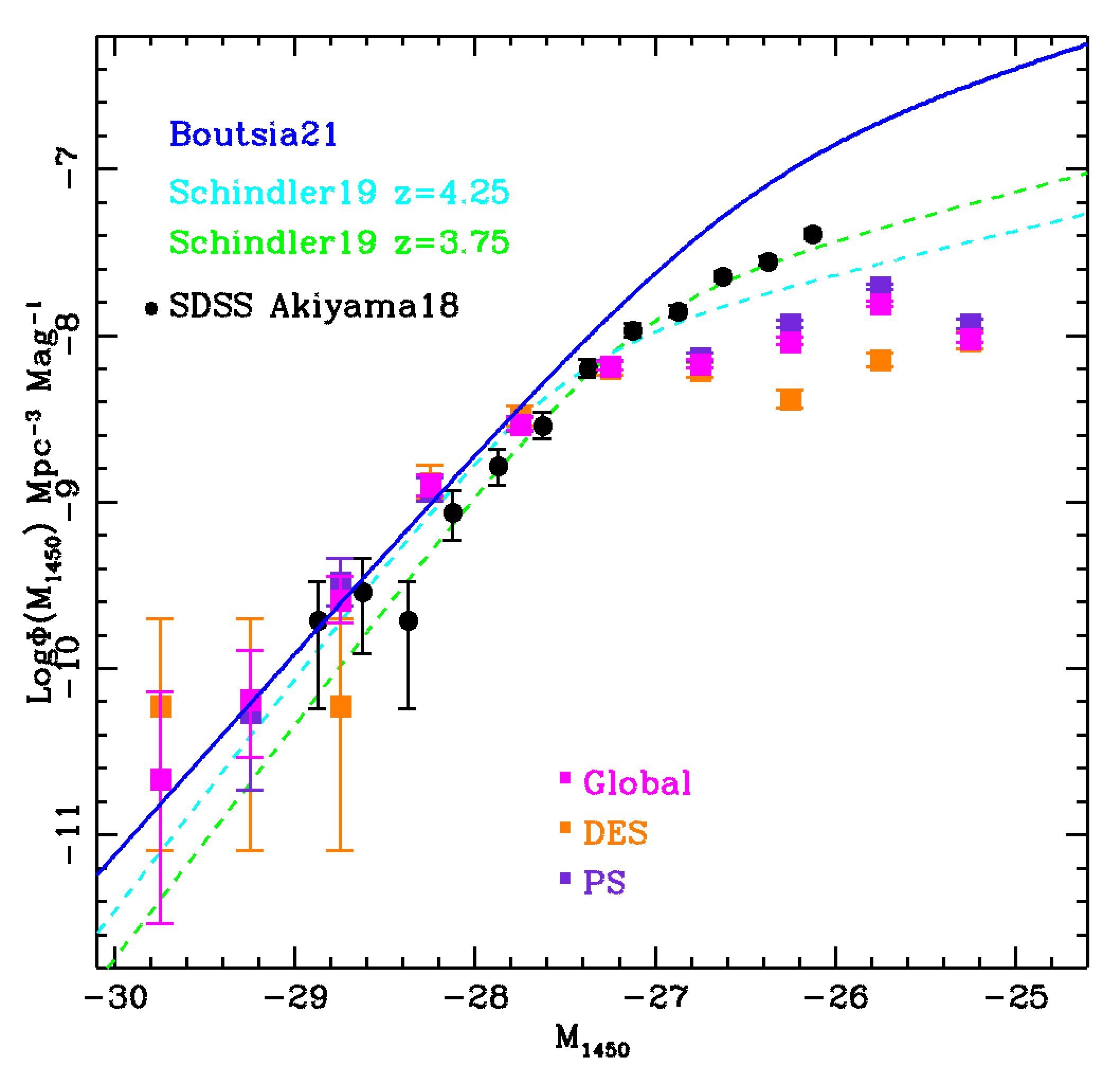}
\caption{The luminosity function of AGNs at z$\sim$4 for the 
QUBRICS-only sample. The orange points are for the DES survey, the violet for PanSTARRS,
and the magenta for the merged sample dubbed as Global. For comparison we show the space densities derived by \cite{Akiyama18} based on SDSS data (black points); the LFs determined by \cite{Sch19a} for two redshifts close to the limits of our redshift range (green and cyan dotted lines); and the LF determined by \cite{boutsia21} (blue line). }
\label{fig:lfsep}
\end{figure}

As mentioned above, we first calculate a separate LF for each dataset, thus
we have a LF based on the DES database and one for the PanSTARRS database. Since the two curves were consistent,
we decided to merge them and calculate a single, global LF for the QUBRICS survey (see Fig.\ref{fig:lfsep}).
As can be seen in Fig.\ref{fig:lfsep}, at magnitudes fainter than M$_{1450}$=-27 our sample is affected by incompleteness. Thus, for the final fit of the bright-end slope we will only consider the QUBRICS points brighter than M$_{1450}$=-27.5. 

The rather steep resulting bright-end slope of the global LF (M$_{1450}<$-27) is consistent with the previous determination of the LF, presented by \cite{boutsia21}. 
Indeed, it corresponds to space densities larger than previous results by SDSS and \cite{Sch19b} at the same redshift, already without correcting for incompleteness. 

\subsection{Completeness of the QUBRICS survey}

We can derive a rough estimate for the completeness thanks to the Gaia DR3 dataset. Considering the entire QUBRICS sample and not just the redshift range between 3.6$<z<$4.2, we conducted a completeness study based on the comparison of our candidates with an independent sample based on the Gaia DR3 low-resolution spectra \citep{Porru26}. The $\sim$3500 QSOs, independently retrieved by Gaia and included in the QUBRICS footprint, were cross-matched with the results of the QUBRICS selection methods to determine how many of the known QSOs were correctly selected as candidates. Based on this analysis, the XGB algorithm reaches a candidate selection completeness of 87\% and the PRF algorithm is slightly less efficient, reaching a 64\% completeness. These values are consistent with the ones originally published in the respective papers \citep{calderone23,guarneri22} and now they have also been validated against independent datasets. Since the sample that we are using in this work is the result of the entire QUBRICS effort to date, the completeness of the QSO subsample presented here is of the order of 82\%, based on the results presented by \cite{Porru26}. In Fig.\ref{fig:lfdesi}, we show the global LF with and without completness correction. The numbers reported in Table \ref{tab:lffit} have been corrected for an average completeness of 82\%.

\subsection{Completeness of the DESI DR1 survey}

\begin{figure}
\includegraphics[width=\linewidth]{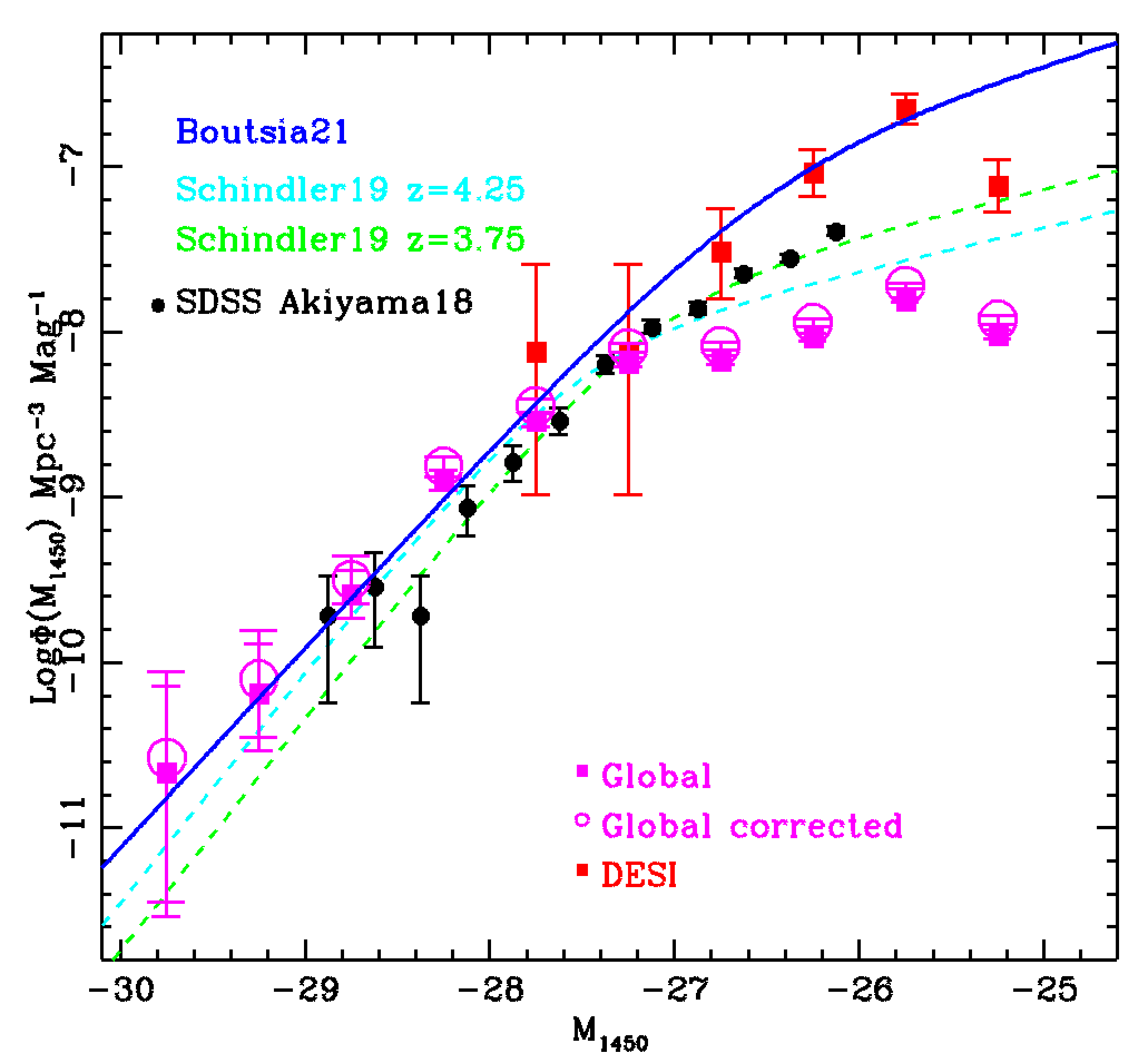}
\caption{The luminosity function of AGNs at z$\sim$4 for the 
QUBRICS and DESI samples. The magenta points indicate the global LF based on the merging of DES and PanSTARRS data and the red points indicate the LF calculated based on the DESI DR1 data. The magenta circles are the QUBRICS points considering an 82\% completeness. The data from the literature are the same as in Fig.\ref{fig:lfsep}. }
\label{fig:lfdesi}
\end{figure}

For the fainter magnitude bins, we used the LF points derived from the DESI sample, after applying the same 
method to calculate the absolute magnitudes as in the rest of the sample (Fig.\ref{fig:lfdesi}). 
Again, in the last bin (M$_{1450}=$-25.25), the DESI LF starts to drop, thus we are affected by incompleteness. 
This last point will not be considered for the final fit. In the same figure, we show for comparison, the QUBRICS LF with and without completeness correction. Since the completeness correction applied to the QUBRICS datapoints is based on the work by \cite{Porru26}, it does not evolve with magnitude. We see that, considering the errors, there is no significant difference between the two sets of points.  

Table \ref{tab:SDqubrics} summarizes the observed space densities of bright QSOs found by QUBRICS in the redshift
interval $3.6\le z\le 4.2$, corrected considering a completeness of 82\%, and Table \ref{tab:SDdesi} presents the space densities based on the DESI data in the same redshift interval. 

In order to derive the completeness of the AGN LF for the DESI DR1 dataset, we compare the QSOs emerged from this survey with the ones already known from the literature, in particular by the QUBRICS compilation \citep{Cristiani23}. A total of 46 QSOs have previously been known in the DESI area by the QUBRICS collection, while only 44 of these have been selected by DESI DR1.
It is interesting to note that the QSOs missing by DESI are mainly in the bright side of the LF, so it could be an indication that the DESI selection criteria could be biased against the brightest QSOs. As we discussed in the previous sub-section, we decided not to correct for incompleteness, since it is not a trivial task. Instead we compute the AGN LF in the DESI area by also adding the QSOs known from QUBRICS but missed by DESI and we consider it as a lower limit. 

\begin{table*}
\caption{The space density $\Phi$ of $3.6\le z\le 4.2$ QSOs in the QUBRICS footprint (DES+PanSTARRS) after applying the correction for an 82\% completeness.}
\label{tab:SDqubrics}
\begin{center}
\begin{tabular}{c c c c c c}
\hline
\hline
Interval & $<M_{1450}>$ & $N_{QSO}$ & $\Phi$ & $\sigma_\Phi(up)$
& $\sigma_\Phi(low)$ \\
 & & & $cMpc^{-3}$ & $cMpc^{-3}$ & $cMpc^{-3}$ \\
\hline
$-28.0\le M_{1450}\le -27.5$ & -27.75 &  135 &  3.578E-09  & 3.353E-10 & 3.077E-10 \\
$-28.5\le M_{1450}\le -28.0$ & -28.25 &   58 &  1.537E-09  & 2.296E-10 & 2.014E-10 \\
$-29.0\le M_{1450}\le -28.5$ & -28.75 &   12 &  3.180E-10  & 1.211E-10 & 9.085E-11 \\
$-29.5\le M_{1450}\le -29.0$ & -29.25 &    3 &  7.951E-11  & 7.783E-11 & 4.395E-11 \\
$-30.0\le M_{1450}\le -29.5$ & -29.75 &    1 &  2.650E-11  & 6.156E-11 & 2.295E-11 \\
\hline
\hline
\end{tabular}
\end{center}
\end{table*}

\begin{table*}
\caption{The space density $\Phi$ of $3.6\le z\le 4.2$ QSOs based on DESI DR1 data.}
\label{tab:SDdesi}
\begin{center}
\begin{tabular}{c c c c c c}
\hline
\hline
Interval & $<M_{1450}>$ & $N_{QSO}$ & $\Phi$ & $\sigma_\Phi(up)$
& $\sigma_\Phi(low)$ \\
 & & & $cMpc^{-3}$ & $cMpc^{-3}$ & $cMpc^{-3}$ \\
\hline
$-26.0\le M_{1450}\le -25.5$ & -25.75 &  29 &  2.366E-07 & 5.266E-08 & 4.375E-08 \\
$-26.5\le M_{1450}\le -26.0$ & -26.25 &  12 &  9.791E-08 & 3.729E-08 & 2.797E-08 \\
$-27.0\le M_{1450}\le -26.5$ & -26.75 &   4 &  3.264E-08 & 2.594E-08 & 1.580E-08 \\
$-27.5\le M_{1450}\le -27.0$ & -27.25 &   1 &  8.159E-09 & 1.895E-08 & 7.066E-09 \\
\hline
\hline
\end{tabular}
\end{center}
\end{table*}


\section{The best-fit luminosity function of $z\sim 4$ AGNs}
\label{sec:bestFit}

\begin{figure}
\includegraphics[width=\linewidth]{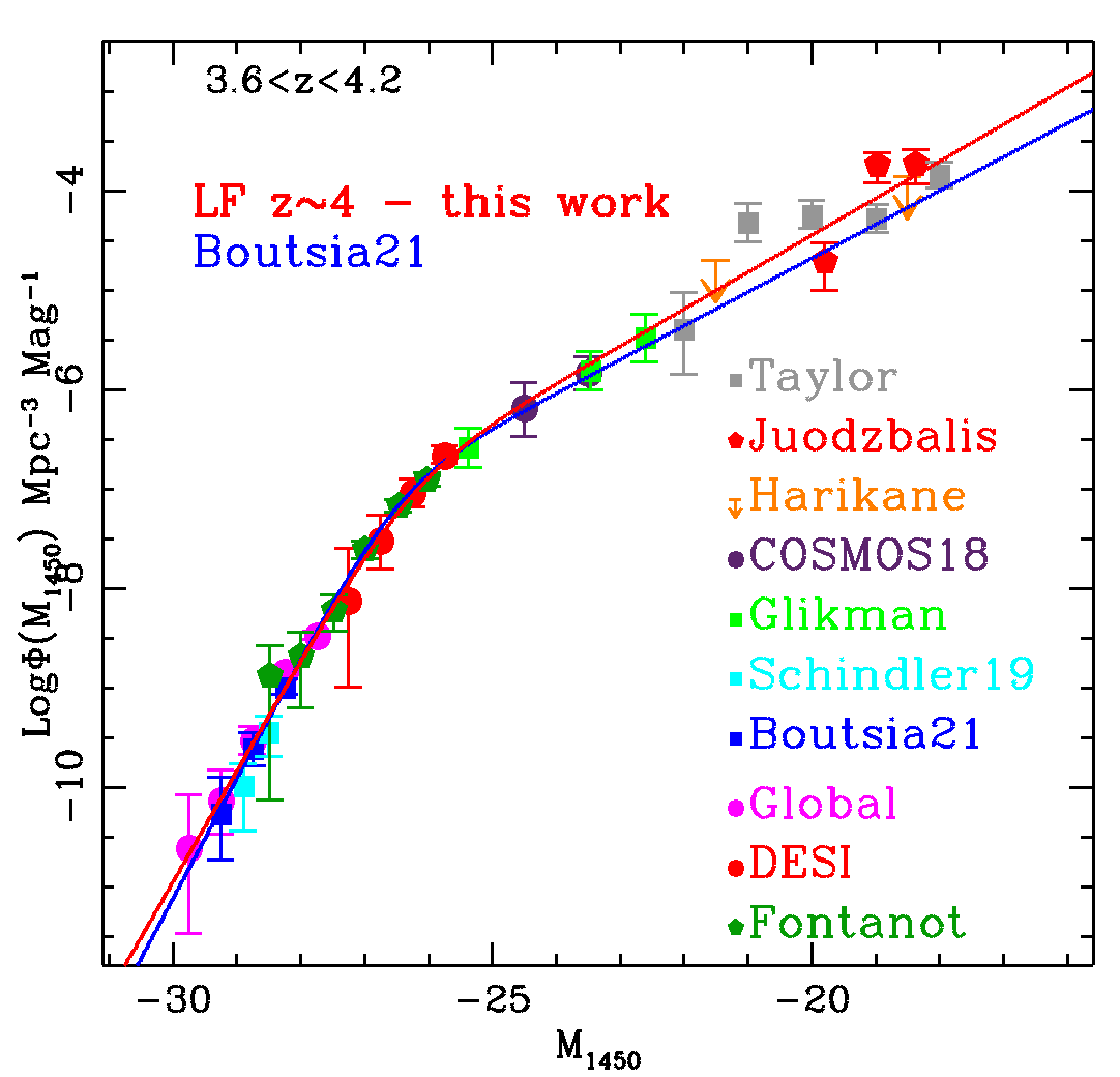}
\caption{The $z\sim 4$ AGN luminosity function. The data points are from QUBRICS and DESI (magenta and red squares, respectively), while the rest of the symbols are from the literature. The red solid line is the best fit to the observed data and the blue dashed line is the fit presented by \cite{boutsia21}.}
\label{fig:lffaint}
\end{figure}

In order to provide a best fit to the QUBRICS and DESI data on a wider magnitude 
range that covers both the bright and faint ends, we also considered 
luminosity function determinations at lower luminosities from the literature.
In our best-fit analysis we included data from \cite{Fontanot07,Glikman11,boutsia18,Sch19a,boutsia21}, that are based on follow-ups from ground-based telescopes. For the faintest bins, we derive our data from recent JWST surveys, including  \citet{Harikane23, Taylor25, Juod26}, as seen in Fig.\ref{fig:lffaint}. For the bright end, we used the data points shown in Tables \ref{tab:SDqubrics} and \ref{tab:SDdesi}.

For this analysis, we adopted a double power-law function for the LF as described below: 

\begin{equation}
    \phi = \frac{\phi*}{10^{0.4(M-M^{*}_{1450})(\alpha+1)}+10^{0.4(M-M^{*}_{1450})(\beta+1)}} \,.
\end{equation}

\begin{table}
\caption{The best fit parameters of the QSO luminosity function at
$3.6\le z\le 4.2$.}
\label{tab:lffit}
\begin{center}
\begin{tabular}{c c c c c c}
\hline
\hline
$\alpha$ & $\beta$ & $M^*_{1450}$ & $Log\Phi^*$ \\
\hline
$-1.93_{-0.08}^{+0.10}$ & $-3.80_{-0.37}^{+0.28}$ & $-26.27_{-0.51}^{+0.63}$
& $ -6.74_{-0.39}^{+0.46}$ \\
\hline
\hline
\end{tabular}
\end{center}
The errors associated to the best fit parameters are at 68\% confidence level (1 $\sigma$).
\end{table}

We define the LF ``knee'' as the break magnitude $M^{*}_{1450}$ of the double power-law form in Eq.~(3), i.e. the magnitude at which the LF transitions from the bright-end slope $\beta$ to the faint-end slope $\alpha$. 

The best fit of the $z\sim 4$ AGN LF
has been carried out by a minimum $\chi^2$ analysis on the above
mentioned binned data points. The best-fit result is shown in Fig. \ref{fig:lffaint} (red line), and the best-fit parameters, together with their 
1$\sigma$ uncertainty ranges are summarized in Table \ref{tab:lffit}. These errors have been derived by imposing a $\Delta\chi^2=\chi^2-\chi^2_{min}\le 4.72$, corresponding to a 68.3\% confidence level for 4 degrees of freedom.

We note that in order to use the volume densities found in the literature at slightly different redshift bins than $z\sim4$, they had to be scaled to our average redshift. To do this, a pure density evolution recipe has been adopted with $\gamma$=-0.25 up to an absolute magnitude of M$_{1450}<$-21, as discussed in \cite{grazian22}. However, in \cite{Grazian24} it was shown that there is a differential evolution between the bright and faint end of the LF, with the faint end remaining almost constant for redshifts $z>4$. For this reason, for sources fainter than M$_{1450}=$-21, we adopted $\gamma$=0, for scaling the volume densities to our average redshift. This mainly affects the data obtained by \cite{Harikane23,Taylor25} and \cite{Juod26}. In Fig. \ref{fig:lffaint} we show the LF based on the sample analyzed in this work (red line) and for comparison we include the best-fit result of \citet{boutsia21} (blue line), where we considered a redshift evolution of $\gamma$=-0.38. At the bright-end the two LFs are practically the same. We see that the current faint-end slope is marginally higher than the one shown by \cite{boutsia21} ($\alpha$=-1.85), however, both fits remain consistent within a 3 $\sigma$ level. In addition, with these new results, the knee of the $z\sim 4$ AGN LF $M^*_{1450}$ is now determined with greater accuracy than in \cite{boutsia21}. 

\section{Comparison with AGN LF at $z\sim 5$ and simulations}
\label{sec:comparison}

\subsection{Comparison with AGN LF at $z\sim 5$}

The most recent derivation of the AGN LF at $z\sim 5$ was reported by \cite{Grazian24}. In that work, three scenarios were explored for the calculation of the LF going from a more to a less conservative one. In Fig.\ref{fig:lfmodel}, we compare the AGN LF at $z\sim4$ with Option 1 and 2 of their work.
Option 1 is the most conservative and includes the space densities based on JWST data, while Option 2 is the intermediate determination \citep[see][for more details]{Grazian24}.

As seen in Fig. \ref{fig:lfmodel} for the interval between $z\sim4$ to $z\sim5$, at the bright-end, the slopes are compatible within a 1-$\sigma$ level. However, the faint-end slope is flatter at $z\sim5$ considering Option 1 where we have $\alpha$=-1.32, compared to $\alpha$=-1.93 at $z\sim4$. This happens, despite the fact that no evolution was considered for computing the faint-end of the LF at $z\sim4$. If a $\gamma$=-0.25 had been adopted, then the faint-end of the LF would have been even steeper. This seems to contradict the assumption made in \cite{Grazian24}, for mild to no evolution in the redshift range 4$<z<$10. Considering Option 2, the faint-end slope at $z\sim5$ is of the order of $\alpha$=-2.22 thus higher than the one at $z\sim4$. In fact, the slope of our LF at $z\sim4$ is within the envelope of the conservative and intermediate estimates for $z\sim5$, leaving room for the no-evolution scenario.
What becomes clear by this analyisis is that the space densities at magnitudes fainter than $M_{1450}$=-19.0 still present significant error bars and further data are necessary to better constrain the faint-end slop. These measurements are currently becoming available thanks to JWST surveys.
 
\subsection{Comparison with theoretical models.}

We compare our observational results with theoretical predictions drawn from  the GAlaxy Evolution and Assembly ({\sc gaea}) semi-analytic model, which features an improved modeling for cold gas accretion onto SMBHs and AGN-driven outﬂows \citep{Fontanot20}. In this work, we consider model predictions from the latest {\sc gaea} implementation \citep{Fontanot25}, run on merger trees extracted from the {\sc  planck millennium} simulation \citep{Baugh19}, which samples the evolution of the matter distribution over a volume box of $\sim 0.5$ Gpc$^3$. The model has been primarly calibrated to reproduce the evolution of the stellar mass function up to z$\sim$3 and on the evolution of the AGN/QSO bolometric LF up to z$\sim$3.5. It is thus worth stressing that values at higher redshifts represent direct predictions of the model. 

In our previous paper \citep{Grazian24}, we already showed that {\sc gaea} was able to correctly reproduce the space density of the $z\sim5$ AGN/QSO population over a wide luminosity range (from M$_{1450} \sim$-28 to M$_{1400} \sim$-17). The main discrepancy between theoretical expectations and the $z\sim5$ data was found around the knee of the LF, that {\sc gaea} predictions tend to underestimate. 

\begin{figure}
\includegraphics[width=\linewidth]{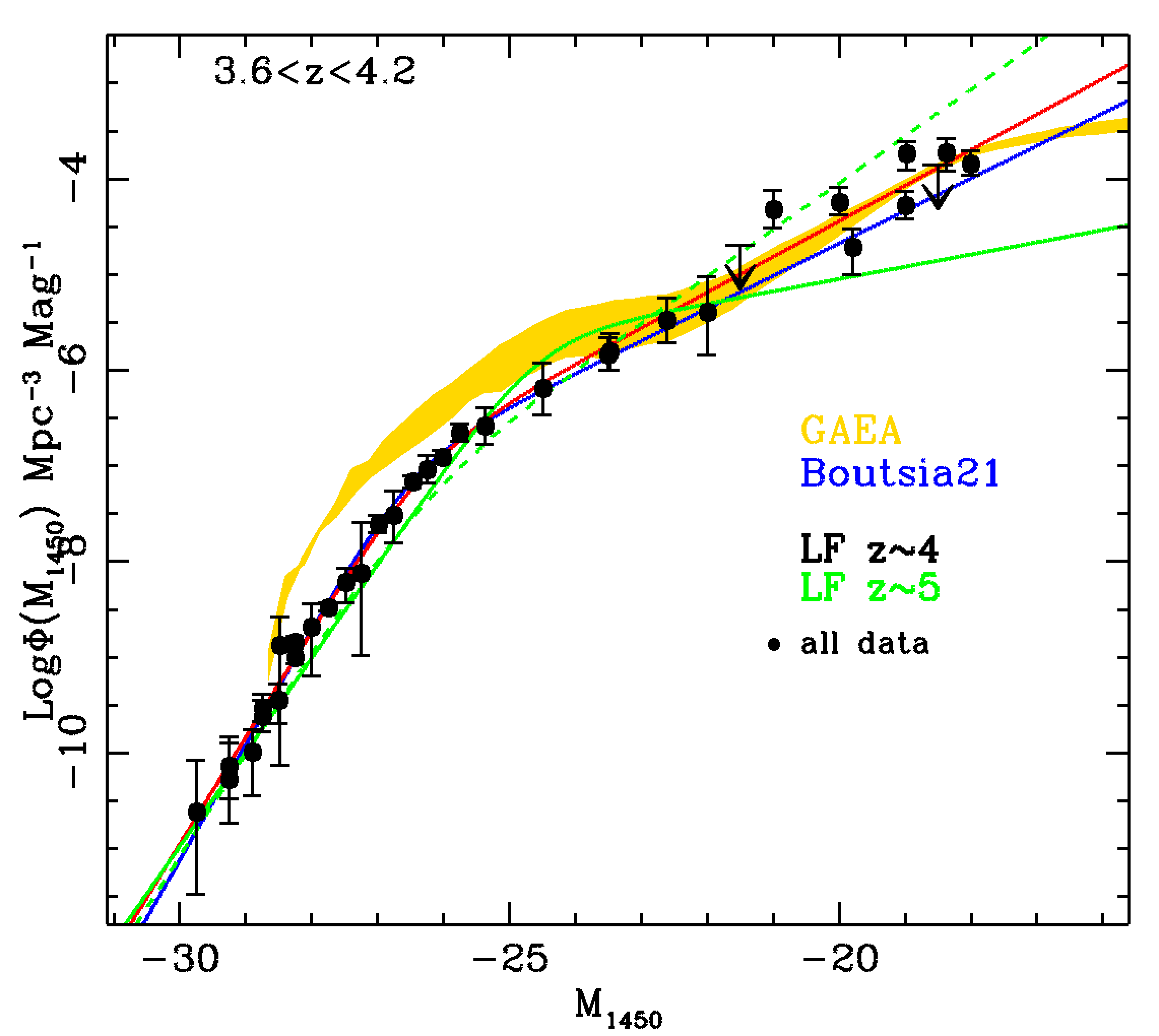}
\caption{The comparison of the $z\sim 4$ AGN LF (red) with the LF at $z\sim 5$ \citep[Option 1 by][ green line, Option 2 green dashed line]{Grazian24} and with numerical predictions at $z\sim 4$ (yellow shaded area defined by model predictions from $z\sim3.6$ to $z\sim4.2$) by the GAEA simulation \citep{Fontanot25}. We also show in blue the previous estimate of the LF at $z\sim 4$ presented in \cite{boutsia21}. The black points are all space densities used to calculate the LF at $z\sim4$ in this work.}
\label{fig:lfmodel}
\end{figure}

In Fig.~\ref{fig:lfmodel}, we compare {\sc gaea} predictions (yellow shaded area) with our proposed fits to observational data. In fact, the yellow shaded area is defined using the range from $z\sim3.6$ to $z\sim4.2$ in model predictions. 
The comparison clearly show that the model tends to overpredict\footnote{Please note that the {\sc gaea} predictions at z$\sim$4 shown in our previous \citet{boutsia21} paper were based on a different (previous) version of the model, using the same prescription for gas accretion onto SMBH, but a different modelling of star formation activity: indeed the latest version of the {\sc gaea} model \citep{DeLucia24} implements an explicit partition of cold gas into atomic and molecular components, with star formation activity tightly linked only to the latter component.} the QSO space density at the bright end of LF. If we compare this result with the z$\sim$5 LF from \citet{Grazian24}, it is clear that the model predicts a much faster evolution of the space density of bright QSOs in the short span between 4$<z<$5 with respect to observational data, which indeed suggest a constant space density in the same range. At magnitudes fainter than M$_{1400} \sim$-22 the model is in agreement with the new estimate proposed in this work. It is worth remembering that the same model is in decent agreement with the faint end at z$\sim$5 \citep{Grazian24}, in the same magnitude range, thus suggesting that the main tension between model and data can be ascribed to a quite fast theoretical evolution of the overall LF at 4$<z<$5, while our observational fits suggest a milder evolution in the same redshift range. Interestingly, a slow evolution of the number density of high-z AGNs can give important constraints on the AGN duty cycle and on the black hole mass-stellar mass scaling relations, as shown by \citet{Roberts26}.

\section{Discussion}
\label{sec:discussion}

In recent years, the estimate of the QSO LF at z$\sim$4 has been revised several times. 
The current dataset represents the state-of-the-art for this subsample of the QSO population and is based on all data available in the literature, including the recent DESI DR1 catalog. The AGN LF shown in Fig. \ref{fig:lffaint} spans a very wide range of more than 10 magnitudes in absolute magnitude $M_{1450}$. Although at faintest magnitudes it could still benefit from more tight constraints from JWST observations, the general picture of the bright side of the AGN LF has not dramatically changed since the work of \cite{boutsia21}. 

\subsection{Absolute magnitudes}
In this work, the absolute magnitudes were calculated based on the Y-band, centered at $\sim$9600\AA, as discussed in Section 3. This choice was motivated by the fact that the selection for the current QUBRICS Main Sample is being conducted in this band. However, in order to ensure that this choice does not significantly bias our results, we have also calculated the LF using the absolute magnitudes in the i-band, which is shown in the Appendix A. As seen in the Table \ref{tab:lffiti}, all derived values for the LF, either based on the i- or Y-band are in agreement within a 1$\sigma$ level. 

\subsection{Completeness }

The selection criteria adopted by the QUBRICS survey have been continuously updated \citep[e.g.][]{calderone19,boutsia20,Guarneri21,guarneri22,calderone23,Cristiani23}, so it is not trivial to define the completeness fraction in a proper way.

An educated guess of the incompleteness still present in our dataset can be derived from the fraction of QSO candidates still lacking spectroscopic identification.
Currently, in the DES footprint there are $\sim$200 QUBRICS candidates with magnitudes brighter than $Y-band$=20.0 to be observed, while there are no candidates with $Y-band<$18.5 still lacking spectroscopic identification. In the PanSTARRS area, instead, there are more than 10,000 candidates at $Y-band\le 20.0$ and only 60 at $Y-band\le 18.5$. In total, we have 60 unidentified candidates at $Y-band\le 18.5$ out of a total of 511 confirmed QSOs at the same magnitude limit. Thus we can consider an incompleteness level of 11.7\%, slightly less but still consistent with the proper estimates of 13-27\% made by \cite{Porru26}. For this reason, we considered an 82\% completeness for the QUBRICS dataset and applied the relative corrections. 

It is interesting to note that the incompleteness we derive in this work ($\sim 10-30\%$) is of the same order of the one derived by \citet{Sch19b} and by \citet{boutsia21}, when compared with previous AGN surveys at $z\sim 4$ \citep[e.g.][]{Aki18}. 

\subsection{The faint side of the AGN LF at $z\sim 4$}

With the advent of the JWST era, faint AGN at $-22<M_{1450}<-18$ and $z>$4 are reported \citep[e.g.][]{Harikane23,maiolino23lf,kocevski24} in substantial numbers compared to what the pre-JWST surveys suggested \citep[e.g.][]{Akiyama18}. In fact, based on these recent works by JWST, the space density at the faint end of the LF can be 1 to 2 dex higher than previous estimates. Reported values for the faint end slope of the $z\sim 4$ AGN LF cluster around $-1.5<\alpha<-2$ but with large field-to-field variation and method-dependent scatter. Although the faint-end is still subject to large uncertainties due to the small survey volumes of deep JWST fields, leading to cosmic variance and small number statistics, it is now clear that the number of QSOs at $z\sim$4 is substantially higher than previously thought. At faint magnitudes and higher redshifts, the results of JWST are still uncertain, see e.g. \citet{Grazian24} at $z=5$, but agree on large space densities of high-z, faint AGNs.
If confirmed, this large faint-AGN population could contribute significantly to the UV ionizing emissivity (and to early black hole growth), but uncertainties in escape fractions, obscuration, and the true AGN fraction remain. Recent modeling papers explore both QSO-assisted and QSO-dominated reionization scenarios using JWST-informed counts \citep{Madau24, Asthana25, Singha25}. Thus, it is important to study this luminosity range in detail. 

This study puts stringent constraints to the "knee" of the AGN luminosity function allowing for a smooth transition between the bright and faint end. In fact the DESI data nicely overlaps with the QUBRICS data at the bright-end and the points by \cite{Glikman11} and \cite{boutsia21} at the faint-end. 
The position of the `knee' (or break) of the LF ($M^*_{1450}$) is now determined at $\pm 0.6$ mag, compared to the $\pm 0.75$ mag uncertainty derived by \citet{boutsia21}. Moreover, the LF break moves from $M^*_{1450}=-26.5$ of \citet{boutsia21} to -26.27 (from our work here). This is extremely important in order to study the redshift evolution of the AGN LF at $z\sim 4$ and beyond, placing useful constraints on the theoretical models, as discussed in Section \ref{sec:comparison}.


\section{Summary and Conclusions}
\label{sec:conclusions}

This paper presents a significantly extended and more complete
determination of the AGN LF at redshift $z\sim 4$
(specifically $3.6 \le z \le 4.2$).

The study aims to compile an LF spanning a
wide range of over 10 magnitudes in absolute magnitude
($M_{1450}=-30.0$ to $M_{1450}=-18$). This is achieved by combining
three main data sets:
\begin{itemize}
\item
QUBRICS Survey (at the bright end of the AGN LF), containing 2,220
bright, high-$z$ confirmed QSOs across $13,966$ deg$^2$ sky area. This
sample constrains the bright end of the LF and is considered
spectroscopically complete down to $M_{1450} \approx -27$ 
(see Table \ref{tab:glob}.)
\item
DESI DR1 (at the intermediate luminosities of AGN LF): data
from the DESI survey were included to extend the LF to fainter magnitudes, down to $M_{1450}=-25$. This has added 55 sources in an area of 37.2 $deg^{2}$ to our sample.
\item
Literature data including recent JWST surveys (at the faint end of the AGN LF): these data 
are incorporated to constrain the faint
end down to $M_{1450}=-18$. These numbers, based on early results, can be considered as lower limits and currently suggest  
substantially higher number densities of faint AGNs (e.g., including Little Red Dots) than
previous estimates.
\end{itemize}

The AGN space density ($\Phi$) is calculated using the $1/V_{\text{max}}$
method and the full LF is fitted with a double power-law function.
The LF fit covers a wide magnitude range of more than 10 magnitudes,
from $M_{1450}=-18$ to $M_{1450}=-30$, corresponding to a factor of
$\sim$100,000 in luminosity. We have thus extended both the bright and
the faint luminosity regimes with respect to \citet{boutsia21}. 
The combined data sets provide the most stringent constraints
to date on the `knee' ($M_{1450}^{*}$) of the AGN LF at $z \sim 4$,
offering a highly accurate description of the transition between the
bright and faint AGN populations.
As shown in the Appendix, the derived LF values based on Y-band are 
robust and in agreement with the ones derived using the i-band that 
is closer to the absolute magnitude $M_{1450}$. 
The bright and faint ends of the AGN LF are consistent with the
\citet{boutsia21} results (see Fig. \ref{fig:lfmodel}),
however, we showed that the faint-end is still affected by small
number statistics and large sample variance, together with open issues about AGN classification.
Future works should concentrate on the detailed study of the faint
side of the AGN LF with JWST, if possible, in order to precisely determine
the population of feebly accreting SMBHs at $z>4$.

These new results are vital for studying the redshift evolution of
AGNs and constraining theoretical models of supermassive black hole seeding
and cosmic reionization evolution. 
The LF of $z\sim 4$ AGN shown in Fig. \ref{fig:lffaint} is more
extended and precisely known than the AGN LF at lower
redshifts. Thus, it can be considered as a reference to study the SMBH
evolution, both in luminosity and in BH mass, at $z>4$, as discussed
in Section \ref{sec:comparison}. The $z\sim 4$ LF can also be
adopted as a benchmark for theoretical models, in order to put strong
constraints on the key physical parameters of semi analytic
simulations. Last but not least, the $z\sim 4$ AGN LF will be fundamental to make
accurate predictions for the expected numbers of high-z, faint AGNs in
future wide area imaging surveys, e.g. Euclid, Vera Rubin LSST, and Roman.




\begin{acknowledgments}

The work of KB is supported by NOIRLab, which is managed by the Association of Universities for Research in Astronomy (AURA) under a cooperative agreement with the U.S. National Science Foundation.\\
We acknowledge the support of the INAF GO/GTO Grant 2023 ``Finding
the Brightest Cosmic Beacons in the Universe with QUBRICS'' (PI Grazian).\\
AT is partly supported by the INFN INDARK and SISSA ERC IDEAS grants.\\
This work has made use of data from the European Space Agency (ESA) mission
{\it Gaia} (\url{https://www.cosmos.esa.int/gaia}), processed by the {\it Gaia}
Data Processing and Analysis Consortium (DPAC,
\url{https://www.cosmos.esa.int/web/gaia/dpac/consortium}). Funding for the DPAC
has been provided by national institutions, in particular the institutions
participating in the {\it Gaia} Multilateral Agreement. \\
This project used public archival data from the Dark Energy Survey (DES). Funding for the DES Projects has been provided by the U.S. Department of Energy, the U.S. National Science Foundation, the Ministry of Science and Education of Spain, the Science and Technology FacilitiesCouncil of the United Kingdom, the Higher Education Funding Council for England, the National Center for Supercomputing Applications at the University of Illinois at Urbana-Champaign, the Kavli Institute of Cosmological Physics at the University of Chicago, the Center for Cosmology and Astro-Particle Physics at the Ohio State University, the Mitchell Institute for Fundamental Physics and Astronomy at Texas A\&M University, Financiadora de Estudos e Projetos, Funda{\c c}{\~a}o Carlos Chagas Filho de Amparo {\`a} Pesquisa do Estado do Rio de Janeiro, Conselho Nacional de Desenvolvimento Cient{\'i}fico e Tecnol{\'o}gico and the Minist{\'e}rio da Ci{\^e}ncia, Tecnologia e Inova{\c c}{\~a}o, the Deutsche Forschungsgemeinschaft, and the Collaborating Institutions in the Dark Energy Survey.
The Collaborating Institutions are Argonne National Laboratory, the University of California at Santa Cruz, the University of Cambridge, Centro de Investigaciones Energ{\'e}ticas, Medioambientales y Tecnol{\'o}gicas-Madrid, the University of Chicago, University College London, the DES-Brazil Consortium, the University of Edinburgh, the Eidgen{\"o}ssische Technische Hochschule (ETH) Z{\"u}rich,  Fermi National Accelerator Laboratory, the University of Illinois at Urbana-Champaign, the Institut de Ci{\`e}ncies de l'Espai (IEEC/CSIC), the Institut de F{\'i}sica d'Altes Energies, Lawrence Berkeley National Laboratory, the Ludwig-Maximilians Universit{\"a}t M{\"u}nchen and the associated Excellence Cluster Universe, the University of Michigan, the National Optical Astronomy Observatory, the University of Nottingham, The Ohio State University, the OzDES Membership Consortium, the University of Pennsylvania, the University of Portsmouth, SLAC National Accelerator Laboratory, Stanford University, the University of Sussex, and Texas A\&M University.\\
Based in part on observations at Cerro Tololo Inter-American Observatory, National Optical Astronomy Observatory, which is operated by the Association of Universities for Research in Astronomy (AURA) under a cooperative agreement with the National Science Foundation.\\
The Pan-STARRS1 Surveys (PS1) and the PS1 public science archive have been made possible through contributions by the Institute for Astronomy, the University of Hawaii, the Pan-STARRS Project Office, the Max-Planck Society and its participating institutes, the Max Planck Institute for Astronomy, Heidelberg and the Max Planck Institute for Extraterrestrial Physics, Garching, The Johns Hopkins University, Durham University, the University of Edinburgh, the Queen's University Belfast, the Harvard-Smithsonian Center for Astrophysics, the Las Cumbres Observatory Global Telescope Network Incorporated, the National Central University of Taiwan, the Space Telescope Science Institute, the National Aeronautics and Space Administration under Grant No. NNX08AR22G issued through the Planetary Science Division of the NASA Science Mission Directorate, the National Science Foundation Grant No. AST–1238877, the University of Maryland, Eotvos Lorand University (ELTE), the Los Alamos National Laboratory, and the Gordon and Betty Moore Foundation.\\
This research used data obtained with the Dark Energy Spectroscopic Instrument (DESI). DESI construction and operations is managed by the Lawrence Berkeley National Laboratory. This material is based upon work supported by the U.S. Department of Energy, Office of Science, Office of High-Energy Physics, under Contract No. DE–AC02–05CH11231, and by the National Energy Research Scientific Computing Center, a DOE Office of Science User Facility under the same contract. Additional support for DESI was provided by the U.S. National Science Foundation (NSF), Division of Astronomical Sciences under Contract No. AST-0950945 to the NSF’s National Optical-Infrared Astronomy Research Laboratory; the Science and Technology Facilities Council of the United Kingdom; the Gordon and Betty Moore Foundation; the Heising-Simons Foundation; the French Alternative Energies and Atomic Energy Commission (CEA); the National Council of Humanities, Science and Technology of Mexico (CONAHCYT); the Ministry of Science and Innovation of Spain (MICINN), and by the DESI Member Institutions: \url{www.desi.lbl.gov/collaborating-institutions}. The DESI collaboration is honored to be permitted to conduct scientific research on I’oligam Du’ag (Kitt Peak), a mountain with particular significance to the Tohono O’odham Nation. Any opinions, findings, and conclusions or recommendations expressed in this material are those of the author(s) and do not necessarily reflect the views of the U.S. National Science Foundation, the U.S. Department of Energy, or any of the listed funding agencies.\\
The LBT is an international collaboration among institutions in the United States and Europe. At the time data were acquired for this research, LBT Corporation Members were the University of Arizona on behalf of the Arizona Board of Regents; Istituto Nazionale di Astrofisica, Italy; LBT Beteiligungsgesellschaft, Germany, representing the Max-Planck Society, the Leibniz Institute for Astrophysics Potsdam, and Heidelberg University; and The Ohio State University, representing The Ohio State University, University of Notre Dame, University of Minnesota, and University of Virginia.  This research used the facilities of the Italian Center for Astronomical Archives (IA2) operated by INAF at the Astronomical Observatory of Trieste. Observations have benefited from the use of ALTA Center (alta.arcetri.inaf.it) forecasts performed with the Astro-Meso-Nh model. Initialization data of the ALTA automatic forecast system come from the General Circulation Model (HRES) of the European Centre for Medium Range Weather Forecasts.\\
Based on observations made with the Italian Telescopio Nazionale Galileo (TNG) operated on the island of La Palma by the Fundación Galileo Galilei of the INAF (Istituto Nazionale di Astrofisica) at the Spanish Observatorio del Roque de los Muchachos of the Instituto de Astrofisica de Canarias.\\
We acknowledge the support from the LBT-Italian Coordination Facility for the execution of observations, data distribution and reduction.
This research used the facilities of the Italian Center for Astronomical Archive (IA2) operated by INAF at the Astronomical Observatory of Trieste.\\
Based in part on observations obtained at the Southern Astrophysical Research (SOAR) telescope, which is a joint project of the Minist\'{e}rio da Ci\^{e}ncia, Tecnologia e Inova\c{c}\~{o}es (MCTI/LNA) do Brasil, the U.S. National Science Foundation NOIRLab, the University of North Carolina at Chapel Hill (UNC), and Michigan State University (MSU).\\
Based in part on observations obtained through the Astronomical Event Observatory Network (AEON), a joint endeavor of the Las Cumbres Observatory and of NSF NOIRLab, which is managed by the Association of Universities for Research in Astronomy (AURA) under a cooperative agreement with the U.S. National Science Foundation. \\
This publication makes use of data products from the Wide-field Infrared Survey Explorer, which is a joint project of the University of California, Los Angeles, and the Jet Propulsion Laboratory/California Institute of Technology, funded by the National Aeronautics and Space Administration.\\
\end{acknowledgments}

\bibliography{References}{}

@article{DESIearly,
doi = {10.3847/1538-3881/ad3217},
url = {https://dx.doi.org/10.3847/1538-3881/ad3217},
year = {2024},
month = {jul},
publisher = {The American Astronomical Society},
volume = {168},
number = {2},
pages = {58},
author = {DESI Collaboration and Adame, A. G. and Aguilar, J. and Ahlen, S. and Alam, S. and Aldering, G. and Alexander, D. M. and Alfarsy, R. and Allende Prieto, C. and Alvarez, M. and Alves, O. and Anand, A. and Andrade-Oliveira, F. and Armengaud, E. and Asorey, J. and Avila, S. and Aviles, A. and Bailey, S. and Balaguera-Antolínez, A. and Ballester, O. and Baltay, C. and Bault, A. and Bautista, J. and Behera, J. and Beltran, S. F. and BenZvi, S. and Beraldo e Silva, L. and Bermejo-Climent, J. R. and Berti, A. and Besuner, R. and Beutler, F. and Bianchi, D. and Blake, C. and Blum, R. and Bolton, A. S. and Brieden, S. and Brodzeller, A. and Brooks, D. and Brown, Z. and Buckley-Geer, E. and Burtin, E. and Cabayol-Garcia, L. and Cai, Z. and Canning, R. and Cardiel-Sas, L. and Carnero Rosell, A. and Castander, F. J. and Cervantes-Cota, J. L. and Chabanier, S. and Chaussidon, E. and Chaves-Montero, J. and Chen, S. and Chen, X. and Chuang, C. and Claybaugh, T. and Cole, S. and Cooper, A. P. and Cuceu, A. and Davis, T. M. and Dawson, K. and de Belsunce, R. and de la Cruz, R. and de la Macorra, A. and Della Costa, J. and de Mattia, A. and Demina, R. and Demirbozan, U. and DeRose, J. and Dey, A. and Dey, B. and Dhungana, G. and Ding, J. and Ding, Z. and Doel, P. and Doshi, R. and Douglass, K. and Edge, A. and Eftekharzadeh, S. and Eisenstein, D. J. and Elliott, A. and Ereza, J. and Escoffier, S. and Fagrelius, P. and Fan, X. and Fanning, K. and Fawcett, V. A. and Ferraro, S. and Flaugher, B. and Font-Ribera, A. and Forero-Romero, J. E. and Forero-Sánchez, D. and Frenk, C. S. and Gänsicke, B. T. and García, L. Á. and García-Bellido, J. and Garcia-Quintero, C. and Garrison, L. H. and Gil-Marín, H. and Golden-Marx, J. and Gontcho A Gontcho, S. and Gonzalez-Morales, A. X. and Gonzalez-Perez, V. and Gordon, C. and Graur, O. and Green, D. and Gruen, D. and Guy, J. and Hadzhiyska, B. and Hahn, C. and Han, J. J. and Hanif, M. M. S and Herrera-Alcantar, H. K. and Honscheid, K. and Hou, J. and Howlett, C. and Huterer, D. and Iršič, V. and Ishak, M. and Jacques, A. and Jana, A. and Jiang, L. and Jimenez, J. and Jing, Y. P. and Joudaki, S. and Joyce, R. and Jullo, E. and Juneau, S. and Karaçaylı, N. G. and Karim, T. and Kehoe, R. and Kent, S. and Khederlarian, A. and Kim, S. and Kirkby, D. and Kisner, T. and Kitaura, F. and Kizhuprakkat, N. and Kneib, J. and Koposov, S. E. and Kovács, A. and Kremin, A. and Krolewski, A. and L’Huillier, B. and Lahav, O. and Lambert, A. and Lamman, C. and Lan, T.-W. and Landriau, M. and Lang, D. and Lange, J. U. and Lasker, J. and Leauthaud, A. and Le Guillou, L. and Levi, M. E. and Li, T. S. and Linder, E. and Lyons, A. and Magneville, C. and Manera, M. and Manser, C. J. and Margala, D. and Martini, P. and McDonald, P. and Medina, G. E. and Medina-Varela, L. and Meisner, A. and Mena-Fernández, J. and Meneses-Rizo, J. and Mezcua, M. and Miquel, R. and Montero-Camacho, P. and Moon, J. and Moore, S. and Moustakas, J. and Mueller, E. and Mundet, J. and Muñoz-Gutiérrez, A. and Myers, A. D. and Nadathur, S. and Napolitano, L. and Neveux, R. and Newman, J. A. and Nie, J. and Nikutta, R. and Niz, G. and Norberg, P. and Noriega, H. E. and Paillas, E. and Palanque-Delabrouille, N. and Palmese, A. and Pan, Z. and Parkinson, D. and Penmetsa, S. and Percival, W. J. and Pérez-Fernández, A. and Pérez-Ràfols, I. and Pieri, M. and Poppett, C. and Porredon, A. and Pothier, S. and Prada, F. and Pucha, R. and Raichoor, A. and Ramírez-Pérez, C. and Ramirez-Solano, S. and Rashkovetskyi, M. and Ravoux, C. and Rocher, A. and Rockosi, C. and Ross, A. J. and Rossi, G. and Ruggeri, R. and Ruhlmann-Kleider, V. and Sabiu, C. G. and Said, K. and Saintonge, A. and Samushia, L. and Sanchez, E. and Saulder, C. and Schaan, E. and Schlafly, E. F. and Schlegel, D. and Scholte, D. and Schubnell, M. and Seo, H. and Shafieloo, A. and Sharples, R. and Sheu, W. and Silber, J. and Sinigaglia, F. and Siudek, M. and Slepian, Z. and Smith, A. and Soumagnac, M. T. and Sprayberry, D. and Stephey, L. and Suárez-Pérez, J. and Sun, Z. and Tan, T. and Tarlé, G. and Tojeiro, R. and Ureña-López, L. A. and Vaisakh, R. and Valcin, D. and Valdes, F. and Valluri, M. and Vargas-Magaña, M. and Variu, A. and Verde, L. and Walther, M. and Wang, B. and Wang, M. S. and Weaver, B. A. and Weaverdyck, N. and Wechsler, R. H. and White, M. and Xie, Y. and Yang, J. and Yèche, C. and Yu, J. and Yuan, S. and Zhang, H. and Zhang, Z. and Zhao, C. and Zheng, Z. and Zhou, R. and Zhou, Z. and Zou, H. and Zou, S. and Zu, Y.},
title = {The Early Data Release of the Dark Energy Spectroscopic Instrument},
journal = {The Astronomical Journal}
}

@ARTICLE{PRF2019,
       author = {{Reis}, Itamar and {Baron}, Dalya and {Shahaf}, Sahar},
        title = "{Probabilistic Random Forest: A Machine Learning Algorithm for Noisy Data Sets}",
      journal = {\aj},
         year = 2019,
        month = jan,
       volume = {157},
       number = {1},
          eid = {16},
        pages = {16},
          doi = {10.3847/1538-3881/aaf101},
archivePrefix = {arXiv},
       eprint = {1811.05994},
 primaryClass = {astro-ph.IM},
       adsurl = {https://ui.adsabs.harvard.edu/abs/2019AJ....157...16R}
}

@ARTICLE{Almeida23,
       author = {{Almeida}, Andr{\'e}s and {Anderson}, Scott F. and {Argudo-Fern{\'a}ndez}, Maria and {Badenes}, Carles and {Barger}, Kat and {Barrera-Ballesteros}, Jorge K. and {Bender}, Chad F. and {Benitez}, Erika and {Besser}, Felipe and {Bird}, Jonathan C. and {Bizyaev}, Dmitry and {Blanton}, Michael R. and {Bochanski}, John and {Bovy}, Jo and {Brandt}, William Nielsen and {Brownstein}, Joel R. and {Buchner}, Johannes and {Bulbul}, Esra and {Burchett}, Joseph N. and {Cano D{\'\i}az}, Mariana and {Carlberg}, Joleen K. and {Casey}, Andrew R. and {Chandra}, Vedant and {Cherinka}, Brian and {Chiappini}, Cristina and {Coker}, Abigail A. and {Comparat}, Johan and {Conroy}, Charlie and {Contardo}, Gabriella and {Cortes}, Arlin and {Covey}, Kevin and {Crane}, Jeffrey D. and {Cunha}, Katia and {Dabbieri}, Collin and {Davidson}, James W. and {Davis}, Megan C. and {de Andrade Queiroz}, Anna Barbara and {De Lee}, Nathan and {M{\'e}ndez Delgado}, Jos{\'e} Eduardo and {Demasi}, Sebastian and {Di Mille}, Francesco and {Donor}, John and {Dow}, Peter and {Dwelly}, Tom and {Eracleous}, Mike and {Eriksen}, Jamey and {Fan}, Xiaohui and {Farr}, Emily and {Frederick}, Sara and {Fries}, Logan and {Frinchaboy}, Peter and {G{\"a}nsicke}, Boris T. and {Ge}, Junqiang and {Gonz{\'a}lez {\'A}vila}, Consuelo and {Grabowski}, Katie and {Grier}, Catherine and {Guiglion}, Guillaume and {Gupta}, Pramod and {Hall}, Patrick and {Hawkins}, Keith and {Hayes}, Christian R. and {Hermes}, J.~J. and {Hern{\'a}ndez-Garc{\'\i}a}, Lorena and {Hogg}, David W. and {Holtzman}, Jon A. and {Ibarra-Medel}, Hector Javier and {Ji}, Alexander and {Jofre}, Paula and {Johnson}, Jennifer A. and {Jones}, Amy M. and {Kinemuchi}, Karen and {Kluge}, Matthias and {Koekemoer}, Anton and {Kollmeier}, Juna A. and {Kounkel}, Marina and {Krishnarao}, Dhanesh and {Krumpe}, Mirko and {Lacerna}, Ivan and {Lago}, Paulo Jakson Assuncao and {Laporte}, Chervin and {Liu}, Chao and {Liu}, Ang and {Liu}, Xin and {Lopes}, Alexandre Roman and {Macktoobian}, Matin and {Majewski}, Steven R. and {Malanushenko}, Viktor and {Maoz}, Dan and {Masseron}, Thomas and {Masters}, Karen L. and {Matijevic}, Gal and {McBride}, Aidan and {Medan}, Ilija and {Merloni}, Andrea and {Morrison}, Sean and {Myers}, Natalie and {M{\'e}sz{\'a}ros}, Szabolcs and {Negrete}, C. Alenka and {Nidever}, David L. and {Nitschelm}, Christian and {Oravetz}, Daniel and {Oravetz}, Audrey and {Pan}, Kaike and {Peng}, Yingjie and {Pinsonneault}, Marc H. and {Pogge}, Rick and {Qiu}, Dan and {Ramirez}, Solange V. and {Rix}, Hans-Walter and {Fern{\'a}ndez Rosso}, Daniela and {Runnoe}, Jessie and {Salvato}, Mara and {Sanchez}, Sebastian F. and {Santana}, Felipe A. and {Saydjari}, Andrew and {Sayres}, Conor and {Schlaufman}, Kevin C. and {Schneider}, Donald P. and {Schwope}, Axel and {Serna}, Javier and {Shen}, Yue and {Sobeck}, Jennifer and {Song}, Ying-Yi and {Souto}, Diogo and {Spoo}, Taylor and {Stassun}, Keivan G. and {Steinmetz}, Matthias and {Straumit}, Ilya and {Stringfellow}, Guy and {S{\'a}nchez-Gallego}, Jos{\'e} and {Taghizadeh-Popp}, Manuchehr and {Tayar}, Jamie and {Thakar}, Ani and {Tissera}, Patricia B. and {Tkachenko}, Andrew and {Hernandez Toledo}, Hector and {Trakhtenbrot}, Benny and {Fern{\'a}ndez-Trincado}, Jos{\'e} G. and {Troup}, Nicholas and {Trump}, Jonathan R. and {Tuttle}, Sarah and {Ulloa}, Natalie and {Vazquez-Mata}, Jose Antonio and {Vera Alfaro}, Pablo and {Villanova}, Sandro and {Wachter}, Stefanie and {Weijmans}, Anne-Marie and {Wheeler}, Adam and {Wilson}, John and {Wojno}, Leigh and {Wolf}, Julien and {Xue}, Xiang-Xiang and {Ybarra}, Jason E. and {Zari}, Eleonora and {Zasowski}, Gail},
        title = "{The Eighteenth Data Release of the Sloan Digital Sky Surveys: Targeting and First Spectra from SDSS-V}",
      journal = {\apjs},
         year = 2023,
        month = aug,
       volume = {267},
       number = {2},
          eid = {44},
        pages = {44},
          doi = {10.3847/1538-4365/acda98},
archivePrefix = {arXiv},
       eprint = {2301.07688},
 primaryClass = {astro-ph.GA},
       adsurl = {https://ui.adsabs.harvard.edu/abs/2023ApJS..267...44A}
}

@ARTICLE{Cupani22,
       author = {{Cupani}, Guido and {Calderone}, Giorgio and {Selvelli}, Pierluigi and {Cristiani}, Stefano and {Boutsia}, Konstantina and {Grazian}, Andrea and {Fontanot}, Fabio and {Guarneri}, Francesco and {D'Odorico}, Valentina and {Giallongo}, Emanuele and {Menci}, Nicola},
        title = "{Near-infrared spectroscopy of extreme BAL QSOs from the QUBRICS bright quasar survey}",
      journal = {\mnras},
         year = 2022,
        month = feb,
       volume = {510},
       number = {2},
        pages = {2509-2528},
          doi = {10.1093/mnras/stab3562},
archivePrefix = {arXiv},
       eprint = {2112.02594},
 primaryClass = {astro-ph.CO},
       adsurl = {https://ui.adsabs.harvard.edu/abs/2022MNRAS.510.2509C}
}

@ARTICLE{Roberts26,
       author = {{Roberts}, Daniel and {Shankar}, Francesco and {Cammelli}, Vieri and {Fontanot}, Fabio and {Trinca}, Alessandro and {Bisigello}, Laura and {Bont{\`a}}, Elena Dalla and {Fu}, Hao and {Gilli}, Roberto and {Grazian}, Andrea and {Graziani}, Luca and {Lapi}, Andrea and {Menci}, Nicola and {Scholtz}, Jan and {Varadarajan}, Karthik Mahesh},
        title = "{Too many or too massive? Investigating the high-z demography of active SMBHs from JWST}",
      journal = {\mnras},
         year = 2026,
        month = mar,
       volume = {546},
       number = {4},
          eid = {stag223},
        pages = {stag223},
          doi = {10.1093/mnras/stag223},
archivePrefix = {arXiv},
       eprint = {2601.23250},
 primaryClass = {astro-ph.GA},
       adsurl = {https://ui.adsabs.harvard.edu/abs/2026MNRAS.546ag223R}
}

@ARTICLE{Singha25,
       author = {{Singha}, Mainak and {Malhotra}, Sangeeta and {Ely Rhoads}, James},
        title = "{Faint active galactic nuclei supplied 31-75\% of hydrogen-ionizing photons at z>5}",
      journal = {arXiv e-prints},
         year = 2025,
        month = oct,
          eid = {arXiv:2510.26990},
        pages = {arXiv:2510.26990},
          doi = {10.48550/arXiv.2510.26990},
archivePrefix = {arXiv},
       eprint = {2510.26990},
 primaryClass = {astro-ph.GA},
       adsurl = {https://ui.adsabs.harvard.edu/abs/2025arXiv251026990S}
}

@ARTICLE{Madau24,
       author = {{Madau}, Piero and {Giallongo}, Emanuele and {Grazian}, Andrea and {Haardt}, Francesco},
        title = "{Cosmic Reionization in the JWST Era: Back to AGNs?}",
      journal = {arXiv e-prints},
         year = 2024,
        month = jun,
          eid = {arXiv:2406.18697},
        pages = {arXiv:2406.18697},
          doi = {10.48550/arXiv.2406.18697},
archivePrefix = {arXiv},
       eprint = {2406.18697},
 primaryClass = {astro-ph.CO},
       adsurl = {https://ui.adsabs.harvard.edu/abs/2024arXiv240618697M}
}

@ARTICLE{Juod26,
       author = {{Juod{\v{z}}balis}, Ignas and {Maiolino}, Roberto and {Baker}, William M. and {Lake}, Emma Curtis and {Scholtz}, Jan and {D'Eugenio}, Francesco and {Trefoloni}, Bartolomeo and {Isobe}, Yuki and {Tacchella}, Sandro and {Bunker}, Andrew J. and {Carniani}, Stefano and {Charlot}, St{\'e}phane and {Jones}, Gareth C. and {Parlanti}, Eleonora and {Perna}, Michele and {Rinaldi}, Pierluigi and {Robertson}, Brant and {{\"U}bler}, Hannah and {Venturi}, Giacomo and {Willott}, Chris},
        title = "{JADES: comprehensive census of broad-line AGN from reionization to cosmic noon revealed by JWST}",
      journal = {\mnras},
         year = 2026,
        month = mar,
       volume = {546},
       number = {3},
          eid = {stag086},
        pages = {stag086},
          doi = {10.1093/mnras/stag086},
archivePrefix = {arXiv},
       eprint = {2504.03551},
 primaryClass = {astro-ph.GA},
       adsurl = {https://ui.adsabs.harvard.edu/abs/2026MNRAS.546ag086J}
}

@techreport{DESI0717,
    author = {Schlegel, David and Schlafly, Edward},
    title = {Definition of the {DESI} Footprint and Tile Centers},
    institution = {Lawrence Berkeley National Laboratory},
    year = {2019},
    month = {sep},
    number = {DESI-0717},
    type = {Technical Note},
    url = {https://faun.rc.fas.harvard.edu/eschlafly/desi/tiling/dr8/note.pdf},
    note = {Maintained in the "footprint" product}
}

@inproceedings{XGBoost,
  author    = {Chen, Tianqi and Guestrin, Carlos},
  title     = {{XGBoost}: A Scalable Tree Boosting System},
  booktitle = {Proceedings of the 22nd ACM SIGKDD International Conference on Knowledge Discovery and Data Mining},
  series    = {KDD '16},
  year      = {2016},
  pages     = {785--794},
  publisher = {ACM},
  address   = {New York, NY, USA},
  doi       = {10.1145/2939672.2939785},
  url       = {https://doi.org/10.1145/2939672.2939785}
}

@ARTICLE{Taylor25,
       author = {{Taylor}, Anthony J. and {Finkelstein}, Steven L. and {Kocevski}, Dale D. and {Jeon}, Junehyoung and {Bromm}, Volker and {Amor{\'\i}n}, Ricardo O. and {Arrabal Haro}, Pablo and {Backhaus}, Bren E. and {Bagley}, Micaela B. and {Banados}, Eduardo and {Bhatawdekar}, Rachana and {Brooks}, Madisyn and {Calabr{\`o}}, Antonello and {Ch{\'a}vez Ortiz}, {\'O}scar A. and {Cheng}, Yingjie and {Cleri}, Nikko J. and {Cole}, Justin W. and {Davis}, Kelcey and {Dickinson}, Mark and {Donnan}, Callum and {Dunlop}, James S. and {Ellis}, Richard S. and {Fern{\'a}ndez}, Vital and {Fontana}, Adriano and {Fujimoto}, Seiji and {Giavalisco}, Mauro and {Grazian}, Andrea and {Guo}, Jingsong and {Hathi}, Nimish P. and {Holwerda}, Benne W. and {Hirschmann}, Michaela and {Inayoshi}, Kohei and {Kartaltepe}, Jeyhan S. and {Khusanova}, Yana and {Koekemoer}, Anton M. and {Kokorev}, Vasily and {Larson}, Rebecca L. and {Leung}, Gene C.~K. and {Lucas}, Ray A. and {McLeod}, Derek J. and {Napolitano}, Lorenzo and {Onoue}, Masafusa and {Pacucci}, Fabio and {Papovich}, Casey and {P{\'e}rez-Gonz{\'a}lez}, Pablo G. and {Pirzkal}, Nor and {Somerville}, Rachel S. and {Trump}, Jonathan R. and {Wilkins}, Stephen M. and {Yung}, L.~Y. Aaron and {Zhang}, Haowen},
        title = "{Broad-line AGNs at 3.5 < z < 6: The Black Hole Mass Function and a Connection with Little Red Dots}",
      journal = {\apj},
         year = 2025,
        month = jun,
       volume = {986},
       number = {2},
          eid = {165},
        pages = {165},
          doi = {10.3847/1538-4357/add15b},
archivePrefix = {arXiv},
       eprint = {2409.06772},
 primaryClass = {astro-ph.GA},
       adsurl = {https://ui.adsabs.harvard.edu/abs/2025ApJ...986..165T}
}

@ARTICLE{Porru26,
       author = {{Porru}, Matteo and {Cristiani}, Stefano and {Guarneri}, Francesco and {Calderone}, Giorgio and {Grazian}, Andrea and {Boutsia}, Konstantina and {Trost}, Andrea and {D'Odorico}, Valentina and {Cupani}, Guido and {Marques}, Catarina M.~J. and {Chiti Tegli}, Francesco and {Fontanot}, Fabio},
        title = "{Estimating the completeness of the QUBRICS Survey with 3501 QSO redshifts from Gaia DR3 spectra}",
      journal = {arXiv e-prints},
         year = 2026,
        month = mar,
          eid = {arXiv:2603.06907},
        pages = {arXiv:2603.06907},
          doi = {10.48550/arXiv.2603.06907},
archivePrefix = {arXiv},
       eprint = {2603.06907},
 primaryClass = {astro-ph.CO},
       adsurl = {https://ui.adsabs.harvard.edu/abs/2026arXiv260306907P}
}

@ARTICLE{LRD,
       author = {{Setton}, David J. and {Greene}, Jenny E. and {de Graaff}, Anna and {Ma}, Yilun and {Leja}, Joel and {Matthee}, Jorryt and {Bezanson}, Rachel and {Boogaard}, Leindert A. and {Cleri}, Nikko J. and {Katz}, Harley and {Labbe}, Ivo and {Maseda}, Michael V. and {McConachie}, Ian and {Miller}, Tim B. and {Price}, Sedona H. and {Suess}, Katherine A. and {van Dokkum}, Pieter and {Wang}, Bingjie and {Weibel}, Andrea and {Whitaker}, Katherine E. and {Williams}, Christina C.},
        title = "{Little Red Dots at an Inflection Point: Ubiquitous V-shaped Turnover Consistently Occurs at the Balmer Limit}",
      journal = {\apj},
         year = 2025,
        month = dec,
       volume = {995},
       number = {1},
          eid = {118},
        pages = {118},
          doi = {10.3847/1538-4357/ae1500},
archivePrefix = {arXiv},
       eprint = {2411.03424},
 primaryClass = {astro-ph.GA},
       adsurl = {https://ui.adsabs.harvard.edu/abs/2025ApJ...995..118S}
}

@ARTICLE{Asthana25,
       author = {{Asthana}, Shikhar and {Haehnelt}, Martin G. and {Kulkarni}, Girish and {Bolton}, James S. and {Gaikwad}, Prakash and {Keating}, Laura C. and {Puchwein}, Ewald},
        title = "{The impact of faint AGN discovered by JWST on reionization}",
      journal = {\mnras},
         year = 2025,
        month = oct,
       volume = {542},
       number = {4},
        pages = {2968-2986},
          doi = {10.1093/mnras/staf1387},
archivePrefix = {arXiv},
       eprint = {2409.15453},
 primaryClass = {astro-ph.GA},
       adsurl = {https://ui.adsabs.harvard.edu/abs/2025MNRAS.542.2968A}
}

@ARTICLE{DeAngeli,
       author = {{De Angeli}, F. and {Weiler}, M. and {Montegriffo}, P. and {Evans}, D.~W. and {Riello}, M. and {Andrae}, R. and {Carrasco}, J.~M. and {Busso}, G. and {Burgess}, P.~W. and {Cacciari}, C. and {Davidson}, M. and {Harrison}, D.~L. and {Hodgkin}, S.~T. and {Jordi}, C. and {Osborne}, P.~J. and {Pancino}, E. and {Altavilla}, G. and {Barstow}, M.~A. and {Bailer-Jones}, C.~A.~L. and {Bellazzini}, M. and {Brown}, A.~G.~A. and {Castellani}, M. and {Cowell}, S. and {Delchambre}, L. and {De Luise}, F. and {Diener}, C. and {Fabricius}, C. and {Fouesneau}, M. and {Fr{\'e}mat}, Y. and {Gilmore}, G. and {Giuffrida}, G. and {Hambly}, N.~C. and {Hidalgo}, S. and {Holland}, G. and {Kostrzewa-Rutkowska}, Z. and {van Leeuwen}, F. and {Lobel}, A. and {Marinoni}, S. and {Miller}, N. and {Pagani}, C. and {Palaversa}, L. and {Piersimoni}, A.~M. and {Pulone}, L. and {Ragaini}, S. and {Rainer}, M. and {Richards}, P.~J. and {Rixon}, G.~T. and {Ruz-Mieres}, D. and {Sanna}, N. and {Sarro}, L.~M. and {Rowell}, N. and {Sordo}, R. and {Walton}, N.~A. and {Yoldas}, A.},
        title = "{Gaia Data Release 3. Processing and validation of BP/RP low-resolution spectral data}",
      journal = {\aap},
         year = 2023,
        month = jun,
       volume = {674},
          eid = {A2},
        pages = {A2},
          doi = {10.1051/0004-6361/202243680},
archivePrefix = {arXiv},
       eprint = {2206.06143},
 primaryClass = {astro-ph.IM},
       adsurl = {https://ui.adsabs.harvard.edu/abs/2023A&A...674A...2D}
}

@ARTICLE{SDSS2009,
       author = {{Abazajian}, Kevork N. and {Adelman-McCarthy}, Jennifer K. and {Ag{\"u}eros}, Marcel A. and {Allam}, Sahar S. and {Allende Prieto}, Carlos and {An}, Deokkeun and {Anderson}, Kurt S.~J. and {Anderson}, Scott F. and {Annis}, James and {Bahcall}, Neta A. and {Bailer-Jones}, C.~A.~L. and {Barentine}, J.~C. and {Bassett}, Bruce A. and {Becker}, Andrew C. and {Beers}, Timothy C. and {Bell}, Eric F. and {Belokurov}, Vasily and {Berlind}, Andreas A. and {Berman}, Eileen F. and {Bernardi}, Mariangela and {Bickerton}, Steven J. and {Bizyaev}, Dmitry and {Blakeslee}, John P. and {Blanton}, Michael R. and {Bochanski}, John J. and {Boroski}, William N. and {Brewington}, Howard J. and {Brinchmann}, Jarle and {Brinkmann}, J. and {Brunner}, Robert J. and {Budav{\'a}ri}, Tam{\'a}s and {Carey}, Larry N. and {Carliles}, Samuel and {Carr}, Michael A. and {Castander}, Francisco J. and {Cinabro}, David and {Connolly}, A.~J. and {Csabai}, Istv{\'a}n and {Cunha}, Carlos E. and {Czarapata}, Paul C. and {Davenport}, James R.~A. and {de Haas}, Ernst and {Dilday}, Ben and {Doi}, Mamoru and {Eisenstein}, Daniel J. and {Evans}, Michael L. and {Evans}, N.~W. and {Fan}, Xiaohui and {Friedman}, Scott D. and {Frieman}, Joshua A. and {Fukugita}, Masataka and {G{\"a}nsicke}, Boris T. and {Gates}, Evalyn and {Gillespie}, Bruce and {Gilmore}, G. and {Gonzalez}, Belinda and {Gonzalez}, Carlos F. and {Grebel}, Eva K. and {Gunn}, James E. and {Gy{\"o}ry}, Zsuzsanna and {Hall}, Patrick B. and {Harding}, Paul and {Harris}, Frederick H. and {Harvanek}, Michael and {Hawley}, Suzanne L. and {Hayes}, Jeffrey J.~E. and {Heckman}, Timothy M. and {Hendry}, John S. and {Hennessy}, Gregory S. and {Hindsley}, Robert B. and {Hoblitt}, J. and {Hogan}, Craig J. and {Hogg}, David W. and {Holtzman}, Jon A. and {Hyde}, Joseph B. and {Ichikawa}, Shin-ichi and {Ichikawa}, Takashi and {Im}, Myungshin and {Ivezi{\'c}}, {\v{Z}}eljko and {Jester}, Sebastian and {Jiang}, Linhua and {Johnson}, Jennifer A. and {Jorgensen}, Anders M. and {Juri{\'c}}, Mario and {Kent}, Stephen M. and {Kessler}, R. and {Kleinman}, S.~J. and {Knapp}, G.~R. and {Konishi}, Kohki and {Kron}, Richard G. and {Krzesinski}, Jurek and {Kuropatkin}, Nikolay and {Lampeitl}, Hubert and {Lebedeva}, Svetlana and {Lee}, Myung Gyoon and {Lee}, Young Sun and {French Leger}, R. and {L{\'e}pine}, S{\'e}bastien and {Li}, Nolan and {Lima}, Marcos and {Lin}, Huan and {Long}, Daniel C. and {Loomis}, Craig P. and {Loveday}, Jon and {Lupton}, Robert H. and {Magnier}, Eugene and {Malanushenko}, Olena and {Malanushenko}, Viktor and {Mandelbaum}, Rachel and {Margon}, Bruce and {Marriner}, John P. and {Mart{\'\i}nez-Delgado}, David and {Matsubara}, Takahiko and {McGehee}, Peregrine M. and {McKay}, Timothy A. and {Meiksin}, Avery and {Morrison}, Heather L. and {Mullally}, Fergal and {Munn}, Jeffrey A. and {Murphy}, Tara and {Nash}, Thomas and {Nebot}, Ada and {Neilsen}, Jr., Eric H. and {Newberg}, Heidi Jo and {Newman}, Peter R. and {Nichol}, Robert C. and {Nicinski}, Tom and {Nieto-Santisteban}, Maria and {Nitta}, Atsuko and {Okamura}, Sadanori and {Oravetz}, Daniel J. and {Ostriker}, Jeremiah P. and {Owen}, Russell and {Padmanabhan}, Nikhil and {Pan}, Kaike and {Park}, Changbom and {Pauls}, George and {Peoples}, Jr., John and {Percival}, Will J. and {Pier}, Jeffrey R. and {Pope}, Adrian C. and {Pourbaix}, Dimitri and {Price}, Paul A. and {Purger}, Norbert and {Quinn}, Thomas and {Raddick}, M. Jordan and {Re Fiorentin}, Paola and {Richards}, Gordon T. and {Richmond}, Michael W. and {Riess}, Adam G. and {Rix}, Hans-Walter and {Rockosi}, Constance M. and {Sako}, Masao and {Schlegel}, David J. and {Schneider}, Donald P. and {Scholz}, Ralf-Dieter and {Schreiber}, Matthias R. and {Schwope}, Axel D. and {Seljak}, Uro{\v{s}} and {Sesar}, Branimir and {Sheldon}, Erin and {Shimasaku}, Kazu and {Sibley}, Valena C. and {Simmons}, A.~E. and {Sivarani}, Thirupathi and {Allyn Smith}, J. and {Smith}, Martin C. and {Smol{\v{c}}i{\'c}}, Vernesa and {Snedden}, Stephanie A. and {Stebbins}, Albert and {Steinmetz}, Matthias and {Stoughton}, Chris and {Strauss}, Michael A. and {SubbaRao}, Mark and {Suto}, Yasushi and {Szalay}, Alexander S. and {Szapudi}, Istv{\'a}n and {Szkody}, Paula and {Tanaka}, Masayuki and {Tegmark}, Max and {Teodoro}, Luis F.~A. and {Thakar}, Aniruddha R. and {Tremonti}, Christy A. and {Tucker}, Douglas L. and {Uomoto}, Alan and {Vanden Berk}, Daniel E. and {Vandenberg}, Jan and {Vidrih}, S. and {Vogeley}, Michael S. and {Voges}, Wolfgang and {Vogt}, Nicole P. and {Wadadekar}, Yogesh and {Watters}, Shannon and {Weinberg}, David H. and {West}, Andrew A. and {White}, Simon D.~M. and {Wilhite}, Brian C. and {Wonders}, Alainna C. and {Yanny}, Brian and {Yocum}, D.~R.},
        title = "{The Seventh Data Release of the Sloan Digital Sky Survey}",
      journal = {\apjs},
         year = 2009,
        month = jun,
       volume = {182},
       number = {2},
        pages = {543-558},
          doi = {10.1088/0067-0049/182/2/543},
archivePrefix = {arXiv},
       eprint = {0812.0649},
 primaryClass = {astro-ph},
       adsurl = {https://ui.adsabs.harvard.edu/abs/2009ApJS..182..543A}
}

@ARTICLE{Cristiani23,
       author = {{Cristiani}, Stefano and {Porru}, Matteo and {Guarneri}, Francesco and {Calderone}, Giorgio and {Boutsia}, Konstantina and {Grazian}, Andrea and {Cupani}, Guido and {D'Odorico}, Valentina and {Fontanot}, Fabio and {Martins}, Carlos J.~A.~P. and {Marques}, Catarina M.~J. and {Maitra}, Soumak and {Trost}, Andrea},
        title = "{Spectroscopy of QUBRICS quasar candidates: 1672 new redshifts and a golden sample for the Sandage test of the redshift drift}",
      journal = {\mnras},
         year = 2023,
        month = jun,
       volume = {522},
       number = {2},
        pages = {2019-2028},
          doi = {10.1093/mnras/stad1007},
archivePrefix = {arXiv},
       eprint = {2304.00362},
 primaryClass = {astro-ph.CO},
       adsurl = {https://ui.adsabs.harvard.edu/abs/2023MNRAS.522.2019C}
}

@ARTICLE{Grazian24,
       author = {{Grazian}, Andrea and {Giallongo}, Emanuele and {Boutsia}, Konstantina and {Cristiani}, Stefano and {Fontanot}, Fabio and {Bischetti}, Manuela and {Bisigello}, Laura and {Bongiorno}, Angela and {Calderone}, Giorgio and {Chiti Tegli}, Francesco and {Cupani}, Guido and {De Lucia}, Gabriella and {D'Odorico}, Valentina and {Feruglio}, Chiara and {Fiore}, Fabrizio and {Gandolfi}, Giovanni and {Girardi}, Giorgia and {Guarneri}, Francesco and {Hirschmann}, Michaela and {Porru}, Matteo and {Rodighiero}, Giulia and {Saccheo}, Ivano and {Simioni}, Matteo and {Trost}, Andrea and {Viitanen}, Akke},
        title = "{What Are the Pillars of Reionization? Revising the AGN Luminosity Function at z {\ensuremath{\sim}} 5}",
      journal = {\apj},
         year = 2024,
        month = oct,
       volume = {974},
       number = {1},
          eid = {84},
        pages = {84},
          doi = {10.3847/1538-4357/ad6980},
archivePrefix = {arXiv},
       eprint = {2407.20861},
 primaryClass = {astro-ph.CO},
       adsurl = {https://ui.adsabs.harvard.edu/abs/2024ApJ...974...84G}
}

@ARTICLE{Kocevski24,
       author = {{Kocevski}, Dale D. and {Finkelstein}, Steven L. and {Barro}, Guillermo and {Taylor}, Anthony J. and {Calabr{\`o}}, Antonello and {Laloux}, Brivael and {Buchner}, Johannes and {Trump}, Jonathan R. and {Leung}, Gene C.~K. and {Yang}, Guang and {Dickinson}, Mark and {P{\'e}rez-Gonz{\'a}lez}, Pablo G. and {Pacucci}, Fabio and {Inayoshi}, Kohei and {Somerville}, Rachel S. and {McGrath}, Elizabeth J. and {Akins}, Hollis B. and {Arrabal Haro}, Pablo and {Bagley}, Micaela B. and {Bowler}, Rebecca A.~A. and {Carnall}, Adam and {Casey}, Caitlin M. and {Cheng}, Yingjie and {Cleri}, Nikko J. and {Costantin}, Luca and {Cullen}, Fergus and {Davis}, Kelcey and {Donnan}, Callum T. and {Dunlop}, James S. and {Ellis}, Richard S. and {Ferguson}, Henry C. and {Fujimoto}, Seiji and {Fontana}, Adriano and {Giavalisco}, Mauro and {Grazian}, Andrea and {Grogin}, Norman A. and {Hathi}, Nimish P. and {Hirschmann}, Michaela and {Huertas-Company}, Marc and {Holwerda}, Benne W. and {Illingworth}, Garth and {Juneau}, St{\'e}phanie and {Kartaltepe}, Jeyhan S. and {Koekemoer}, Anton M. and {Li}, Wenxiu and {Lucas}, Ray A. and {Magee}, Dan and {Mason}, Charlotte and {McLeod}, Derek J. and {McLure}, Ross J. and {Napolitano}, Lorenzo and {Papovich}, Casey and {Pirzkal}, Nor and {Rodighiero}, Giulia and {Santini}, Paola and {Wilkins}, Stephen M. and {Yung}, L.~Y. Aaron},
        title = "{The Rise of Faint, Red AGN at $z>4$: A Sample of Little Red Dots in the JWST Extragalactic Legacy Fields}",
      journal = {arXiv e-prints},
         year = 2024,
        month = apr,
          eid = {arXiv:2404.03576},
        pages = {arXiv:2404.03576},
archivePrefix = {arXiv},
       eprint = {2404.03576},
 primaryClass = {astro-ph.GA},
       adsurl = {https://ui.adsabs.harvard.edu/abs/2024arXiv240403576K}
}

@ARTICLE{dayal24,
       author = {{Dayal}, Pratika and {Volonteri}, Marta and {Greene}, Jenny E. and {Kokorev}, Vasily and {Goulding}, Andy D. and {Williams}, Christina C. and {Furtak}, Lukas J. and {Zitrin}, Adi and {Atek}, Hakim and {Chemerynska}, Iryna and {Feldmann}, Robert and {Glazebrook}, Karl and {Labbe}, Ivo and {Nanayakkara}, Themiya and {Oesch}, Pascal A. and {Weaver}, John R.},
        title = "{UNCOVERing the contribution of black holes to reionization in the JWST era}",
      journal = {arXiv e-prints},
         year = 2024,
        month = jan,
          eid = {arXiv:2401.11242},
        pages = {arXiv:2401.11242},
          doi = {10.48550/arXiv.2401.11242},
archivePrefix = {arXiv},
       eprint = {2401.11242},
 primaryClass = {astro-ph.GA},
       adsurl = {https://ui.adsabs.harvard.edu/abs/2024arXiv240111242D}
}

@ARTICLE{DeLucia24,
       author = {{De Lucia}, Gabriella and {Fontanot}, Fabio and {Xie}, Lizhi and {Hirschmann}, Michaela},
        title = "{Tracing the quenching journey across cosmic time}",
      journal = {\aap},
         year = 2024,
        month = jul,
       volume = {687},
          eid = {A68},
        pages = {A68},
          doi = {10.1051/0004-6361/202349045},
archivePrefix = {arXiv},
       eprint = {2401.06211},
 primaryClass = {astro-ph.GA},
       adsurl = {https://ui.adsabs.harvard.edu/abs/2024A&A...687A..68D}
}

@ARTICLE{Fontanot25,
       author = {{Fontanot}, Fabio and {De Lucia}, Gabriella and {Xie}, Lizhi and {Hirschmann}, Michaela and {Baugh}, Carlton and {Helly}, John C.},
        title = "{Galaxy assembly and evolution in the P-Millennium simulation: Galaxy clustering}",
      journal = {\aap},
         year = 2025,
        month = jul,
       volume = {699},
          eid = {A108},
        pages = {A108},
          doi = {10.1051/0004-6361/202452029},
archivePrefix = {arXiv},
       eprint = {2409.02194},
 primaryClass = {astro-ph.GA},
       adsurl = {https://ui.adsabs.harvard.edu/abs/2025A&A...699A.108F}
}

@ARTICLE{Baugh19,
       author = {{Baugh}, C.~M. and {Gonzalez-Perez}, Violeta and {Lagos}, Claudia D.~P. and {Lacey}, Cedric G. and {Helly}, John C. and {Jenkins}, Adrian and {Frenk}, Carlos S. and {Benson}, Andrew J. and {Bower}, Richard G. and {Cole}, Shaun},
        title = "{Galaxy formation in the Planck Millennium: the atomic hydrogen content of dark matter haloes}",
      journal = {\mnras},
         year = 2019,
        month = mar,
       volume = {483},
       number = {4},
        pages = {4922-4937},
          doi = {10.1093/mnras/sty3427},
archivePrefix = {arXiv},
       eprint = {1808.08276},
 primaryClass = {astro-ph.GA},
       adsurl = {https://ui.adsabs.harvard.edu/abs/2019MNRAS.483.4922B}
}

@ARTICLE{Dodorico23,
       author = {{D'Odorico}, Valentina and {Ba{\~n}ados}, E. and {Becker}, G.~D. and {Bischetti}, M. and {Bosman}, S.~E.~I. and {Cupani}, G. and {Davies}, R. and {Farina}, E.~P. and {Ferrara}, A. and {Feruglio}, C. and {Mazzucchelli}, C. and {Ryan-Weber}, E. and {Schindler}, J. -T. and {Sodini}, A. and {Venemans}, B.~P. and {Walter}, F. and {Chen}, H. and {Lai}, S. and {Zhu}, Y. and {Bian}, F. and {Campo}, S. and {Carniani}, S. and {Cristiani}, S. and {Davies}, F. and {Decarli}, R. and {Drake}, A. and {Eilers}, A. -C. and {Fan}, X. and {Gaikwad}, P. and {Gallerani}, S. and {Greig}, B. and {Haehnelt}, M.~G. and {Hennawi}, J. and {Keating}, L. and {Kulkarni}, G. and {Mesinger}, A. and {Meyer}, R.~A. and {Neeleman}, M. and {Onoue}, M. and {Pallottini}, A. and {Qin}, Y. and {Rojas-Ruiz}, S. and {Satyavolu}, S. and {Sebastian}, A. and {Tripodi}, R. and {Wang}, F. and {Wolfson}, M. and {Yang}, J. and {Zanchettin}, M.~V.},
        title = "{XQR-30: The ultimate XSHOOTER quasar sample at the reionization epoch}",
      journal = {\mnras},
         year = 2023,
        month = jul,
       volume = {523},
       number = {1},
        pages = {1399-1420},
          doi = {10.1093/mnras/stad1468},
archivePrefix = {arXiv},
       eprint = {2305.05053},
 primaryClass = {astro-ph.GA},
       adsurl = {https://ui.adsabs.harvard.edu/abs/2023MNRAS.523.1399D}
}

@ARTICLE{volonteri23,
       author = {{Volonteri}, Marta and {Habouzit}, M{\'e}lanie and {Colpi}, Monica},
        title = "{What if young z > 9 JWST galaxies hosted massive black holes?}",
      journal = {\mnras},
         year = 2023,
        month = may,
       volume = {521},
       number = {1},
        pages = {241-250},
          doi = {10.1093/mnras/stad499},
archivePrefix = {arXiv},
       eprint = {2212.04710},
 primaryClass = {astro-ph.GA},
       adsurl = {https://ui.adsabs.harvard.edu/abs/2023MNRAS.521..241V}
}

@ARTICLE{Harikane23,
       author = {{Harikane}, Yuichi and {Zhang}, Yechi and {Nakajima}, Kimihiko and {Ouchi}, Masami and {Isobe}, Yuki and {Ono}, Yoshiaki and {Hatano}, Shun and {Xu}, Yi and {Umeda}, Hiroya},
        title = "{A JWST/NIRSpec First Census of Broad-line AGNs at z = 4-7: Detection of 10 Faint AGNs with M $_{BH}$ {}10$^{6}$-{}10$^{8}$ M $_{{\ensuremath{\odot}}}$ and Their Host Galaxy Properties}",
      journal = {\apj},
         year = 2023,
        month = dec,
       volume = {959},
       number = {1},
          eid = {39},
        pages = {39},
          doi = {10.3847/1538-4357/ad029e},
archivePrefix = {arXiv},
       eprint = {2303.11946},
 primaryClass = {astro-ph.GA},
       adsurl = {https://ui.adsabs.harvard.edu/abs/2023ApJ...959...39H}
}

@ARTICLE{maiolino23lf,
       author = {{Maiolino}, Roberto and {Scholtz}, Jan and {Curtis-Lake}, Emma and {Carniani}, Stefano and {Baker}, William and {de Graaff}, Anna and {Tacchella}, Sandro and {{\"U}bler}, Hannah and {D'Eugenio}, Francesco and {Witstok}, Joris and {Curti}, Mirko and {Arribas}, Santiago and {Bunker}, Andrew J. and {Charlot}, St{\'e}phane and {Chevallard}, Jacopo and {Eisenstein}, Daniel J. and {Egami}, Eiichi and {Ji}, Zhiyuan and {Jones}, Gareth C. and {Lyu}, Jianwei and {Rawle}, Tim and {Robertson}, Brant and {Rujopakarn}, Wiphu and {Perna}, Michele and {Sun}, Fengwu and {Venturi}, Giacomo and {Williams}, Christina C. and {Willott}, Chris},
        title = "{JADES. The diverse population of infant Black Holes at 4<z<11: merging, tiny, poor, but mighty}",
      journal = {arXiv e-prints},
         year = 2023,
        month = aug,
          eid = {arXiv:2308.01230},
        pages = {arXiv:2308.01230},
          doi = {10.48550/arXiv.2308.01230},
archivePrefix = {arXiv},
       eprint = {2308.01230},
 primaryClass = {astro-ph.GA},
       adsurl = {https://ui.adsabs.harvard.edu/abs/2023arXiv230801230M}
}

@ARTICLE{Matsuoka23,
       author = {{Matsuoka}, Yoshiki and {Onoue}, Masafusa and {Iwasawa}, Kazushi and {Strauss}, Michael A. and {Kashikawa}, Nobunari and {Izumi}, Takuma and {Nagao}, Tohru and {Imanishi}, Masatoshi and {Akiyama}, Masayuki and {Silverman}, John D. and {Asami}, Naoko and {Bosch}, James and {Furusawa}, Hisanori and {Goto}, Tomotsugu and {Gunn}, James E. and {Harikane}, Yuichi and {Ikeda}, Hiroyuki and {Inayoshi}, Kohei and {Ishimoto}, Rikako and {Kawaguchi}, Toshihiro and {Kikuta}, Satoshi and {Kohno}, Kotaro and {Komiyama}, Yutaka and {Lee}, Chien-Hsiu and {Lupton}, Robert H. and {Minezaki}, Takeo and {Miyazaki}, Satoshi and {Murayama}, Hitoshi and {Nishizawa}, Atsushi J. and {Oguri}, Masamune and {Ono}, Yoshiaki and {Oogi}, Taira and {Ouchi}, Masami and {Price}, Paul A. and {Sameshima}, Hiroaki and {Sugiyama}, Naoshi and {Tait}, Philip J. and {Takada}, Masahiro and {Takahashi}, Ayumi and {Takata}, Tadafumi and {Tanaka}, Masayuki and {Toba}, Yoshiki and {Wang}, Shiang-Yu and {Yamashita}, Takuji},
        title = "{Quasar Luminosity Function at z = 7}",
      journal = {\apjl},
         year = 2023,
        month = jun,
       volume = {949},
       number = {2},
          eid = {L42},
        pages = {L42},
          doi = {10.3847/2041-8213/acd69f},
archivePrefix = {arXiv},
       eprint = {2305.11225},
 primaryClass = {astro-ph.GA},
       adsurl = {https://ui.adsabs.harvard.edu/abs/2023ApJ...949L..42M}
}

@ARTICLE{calderone23,
       author = {{Calderone}, Giorgio and {Guarneri}, Francesco and {Porru}, Matteo and {Cristiani}, Stefano and {Grazian}, Andrea and {Nicastro}, Luciano and {Bischetti}, Manuela and {Boutsia}, Konstantina and {Cupani}, Guido and {D'Odorico}, Valentina and {Feruglio}, Chiara and {Fontanot}, Fabio},
        title = "{Boost recall in quasi-stellar object selection from highly imbalanced photometric datasets. The reverse selection method}",
      journal = {\aap},
         year = 2024,
        month = mar,
       volume = {683},
          eid = {A34},
        pages = {A34},
          doi = {10.1051/0004-6361/202346625},
archivePrefix = {arXiv},
       eprint = {2312.13194},
 primaryClass = {astro-ph.IM},
       adsurl = {https://ui.adsabs.harvard.edu/abs/2024A&A...683A..34C}
}

@ARTICLE{guarneri22,
       author = {{Guarneri}, Francesco and {Calderone}, Giorgio and {Cristiani}, Stefano and {Porru}, Matteo and {Fontanot}, Fabio and {Boutsia}, Konstantina and {Cupani}, Guido and {Grazian}, Andrea and {D'Odorico}, Valentina and {Murphy}, Michael T. and {Bongiorno}, Angela and {Saccheo}, Ivano and {Nicastro}, Luciano},
        title = "{The probabilistic random forest applied to the QUBRICS survey: improving the selection of high-redshift quasars with synthetic data}",
      journal = {\mnras},
         year = 2022,
        month = dec,
       volume = {517},
       number = {2},
        pages = {2436-2453},
          doi = {10.1093/mnras/stac2733},
archivePrefix = {arXiv},
       eprint = {2209.07257},
 primaryClass = {astro-ph.IM},
       adsurl = {https://ui.adsabs.harvard.edu/abs/2022MNRAS.517.2436G}
}

@ARTICLE{trinca22,
       author = {{Trinca}, Alessandro and {Schneider}, Raffaella and {Valiante}, Rosa and {Graziani}, Luca and {Zappacosta}, Luca and {Shankar}, Francesco},
        title = "{The low-end of the black hole mass function at cosmic dawn}",
      journal = {\mnras},
         year = 2022,
        month = mar,
       volume = {511},
       number = {1},
        pages = {616-640},
          doi = {10.1093/mnras/stac062},
archivePrefix = {arXiv},
       eprint = {2201.02630},
 primaryClass = {astro-ph.GA},
       adsurl = {https://ui.adsabs.harvard.edu/abs/2022MNRAS.511..616T}
}

@ARTICLE{kocevski23,
       author = {{Kocevski}, Dale D. and {Onoue}, Masafusa and {Inayoshi}, Kohei and {Trump}, Jonathan R. and {Arrabal Haro}, Pablo and {Grazian}, Andrea and {Dickinson}, Mark and {Finkelstein}, Steven L. and {Kartaltepe}, Jeyhan S. and {Hirschmann}, Michaela and {Aird}, James and {Holwerda}, Benne W. and {Fujimoto}, Seiji and {Juneau}, St{\'e}phanie and {Amor{\'\i}n}, Ricardo O. and {Backhaus}, Bren E. and {Bagley}, Micaela B. and {Barro}, Guillermo and {Bell}, Eric F. and {Bisigello}, Laura and {Calabr{\`o}}, Antonello and {Cleri}, Nikko J. and {Cooper}, M.~C. and {Ding}, Xuheng and {Grogin}, Norman A. and {Ho}, Luis C. and {Hutchison}, Taylor A. and {Inoue}, Akio K. and {Jiang}, Linhua and {Jones}, Brenda and {Koekemoer}, Anton M. and {Li}, Wenxiu and {Li}, Zhengrong and {McGrath}, Elizabeth J. and {Molina}, Juan and {Papovich}, Casey and {P{\'e}rez-Gonz{\'a}lez}, Pablo G. and {Pirzkal}, Nor and {Wilkins}, Stephen M. and {Yang}, Guang and {Yung}, L.~Y. Aaron},
        title = "{Hidden Little Monsters: Spectroscopic Identification of Low-mass, Broad-line AGNs at z > 5 with CEERS}",
      journal = {\apjl},
         year = 2023,
        month = sep,
       volume = {954},
       number = {1},
          eid = {L4},
        pages = {L4},
          doi = {10.3847/2041-8213/ace5a0},
archivePrefix = {arXiv},
       eprint = {2302.00012},
 primaryClass = {astro-ph.GA},
       adsurl = {https://ui.adsabs.harvard.edu/abs/2023ApJ...954L...4K}
}

@ARTICLE{fontanot23,
       author = {{Fontanot}, Fabio and {Cristiani}, Stefano and {Grazian}, Andrea and {Haardt}, Francesco and {D'Odorico}, Valentina and {Boutsia}, Konstantina and {Calderone}, Giorgio and {Cupani}, Guido and {Guarneri}, Francesco and {Fiorin}, Chiara and {Rodighiero}, Giulia},
        title = "{Eddington accreting black holes in the epoch of reionization}",
      journal = {\mnras},
         year = 2023,
        month = mar,
       volume = {520},
       number = {1},
        pages = {740-749},
          doi = {10.1093/mnras/stad189},
archivePrefix = {arXiv},
       eprint = {2301.07129},
 primaryClass = {astro-ph.CO},
       adsurl = {https://ui.adsabs.harvard.edu/abs/2023MNRAS.520..740F}
}

@ARTICLE{Rakshit20,
       author = {{Rakshit}, Suvendu and {Stalin}, C.~S. and {Kotilainen}, Jari},
        title = "{Spectral Properties of Quasars from Sloan Digital Sky Survey Data Release 14: The Catalog}",
      journal = {\apjs},
         year = 2020,
        month = jul,
       volume = {249},
       number = {1},
          eid = {17},
        pages = {17},
          doi = {10.3847/1538-4365/ab99c5},
archivePrefix = {arXiv},
       eprint = {1910.10395},
 primaryClass = {astro-ph.GA},
       adsurl = {https://ui.adsabs.harvard.edu/abs/2020ApJS..249...17R}
}

@ARTICLE{EDR2024,
       author = {{DESI Collaboration} and {Adame}, A.~G. and {Aguilar}, J. and {Ahlen}, S. and {Alam}, S. and {Aldering}, G. and {Alexander}, D.~M. and {Alfarsy}, R. and {Allende Prieto}, C. and {Alvarez}, M. and {Alves}, O. and {Anand}, A. and {Andrade-Oliveira}, F. and {Armengaud}, E. and {Asorey}, J. and {Avila}, S. and {Aviles}, A. and {Bailey}, S. and {Balaguera-Antol{\'\i}nez}, A. and {Ballester}, O. and {Baltay}, C. and {Bault}, A. and {Bautista}, J. and {Behera}, J. and {Beltran}, S.~F. and {BenZvi}, S. and {Beraldo e Silva}, L. and {Bermejo-Climent}, J.~R. and {Berti}, A. and {Besuner}, R. and {Beutler}, F. and {Bianchi}, D. and {Blake}, C. and {Blum}, R. and {Bolton}, A.~S. and {Brieden}, S. and {Brodzeller}, A. and {Brooks}, D. and {Brown}, Z. and {Buckley-Geer}, E. and {Burtin}, E. and {Cabayol-Garcia}, L. and {Cai}, Z. and {Canning}, R. and {Cardiel-Sas}, L. and {Carnero Rosell}, A. and {Castander}, F.~J. and {Cervantes-Cota}, J.~L. and {Chabanier}, S. and {Chaussidon}, E. and {Chaves-Montero}, J. and {Chen}, S. and {Chen}, X. and {Chuang}, C. and {Claybaugh}, T. and {Cole}, S. and {Cooper}, A.~P. and {Cuceu}, A. and {Davis}, T.~M. and {Dawson}, K. and {de Belsunce}, R. and {de la Cruz}, R. and {de la Macorra}, A. and {Della Costa}, J. and {de Mattia}, A. and {Demina}, R. and {Demirbozan}, U. and {DeRose}, J. and {Dey}, A. and {Dey}, B. and {Dhungana}, G. and {Ding}, J. and {Ding}, Z. and {Doel}, P. and {Doshi}, R. and {Douglass}, K. and {Edge}, A. and {Eftekharzadeh}, S. and {Eisenstein}, D.~J. and {Elliott}, A. and {Ereza}, J. and {Escoffier}, S. and {Fagrelius}, P. and {Fan}, X. and {Fanning}, K. and {Fawcett}, V.~A. and {Ferraro}, S. and {Flaugher}, B. and {Font-Ribera}, A. and {Forero-Romero}, J.~E. and {Forero-S{\'a}nchez}, D. and {Frenk}, C.~S. and {G{\"a}nsicke}, B.~T. and {Garc{\'\i}a}, L. {\'A}. and {Garc{\'\i}a-Bellido}, J. and {Garcia-Quintero}, C. and {Garrison}, L.~H. and {Gil-Mar{\'\i}n}, H. and {Golden-Marx}, J. and {Gontcho A Gontcho}, S. and {Gonzalez-Morales}, A.~X. and {Gonzalez-Perez}, V. and {Gordon}, C. and {Graur}, O. and {Green}, D. and {Gruen}, D. and {Guy}, J. and {Hadzhiyska}, B. and {Hahn}, C. and {Han}, J.~J. and {Hanif}, M.~M.~S. and {Herrera-Alcantar}, H.~K. and {Honscheid}, K. and {Hou}, J. and {Howlett}, C. and {Huterer}, D. and {Ir{\v{s}}i{\v{c}}}, V. and {Ishak}, M. and {Jacques}, A. and {Jana}, A. and {Jiang}, L. and {Jimenez}, J. and {Jing}, Y.~P. and {Joudaki}, S. and {Joyce}, R. and {Jullo}, E. and {Juneau}, S. and {Kara{\c{c}}ayl{\i}}, N.~G. and {Karim}, T. and {Kehoe}, R. and {Kent}, S. and {Khederlarian}, A. and {Kim}, S. and {Kirkby}, D. and {Kisner}, T. and {Kitaura}, F. and {Kizhuprakkat}, N. and {Kneib}, J. and {Koposov}, S.~E. and {Kov{\'a}cs}, A. and {Kremin}, A. and {Krolewski}, A. and {L'Huillier}, B. and {Lahav}, O. and {Lambert}, A. and {Lamman}, C. and {Lan}, T.-W. and {Landriau}, M. and {Lang}, D. and {Lange}, J.~U. and {Lasker}, J. and {Leauthaud}, A. and {Le Guillou}, L. and {Levi}, M.~E. and {Li}, T.~S. and {Linder}, E. and {Lyons}, A. and {Magneville}, C. and {Manera}, M. and {Manser}, C.~J. and {Margala}, D. and {Martini}, P. and {McDonald}, P. and {Medina}, G.~E. and {Medina-Varela}, L. and {Meisner}, A. and {Mena-Fern{\'a}ndez}, J. and {Meneses-Rizo}, J. and {Mezcua}, M. and {Miquel}, R. and {Montero-Camacho}, P. and {Moon}, J. and {Moore}, S. and {Moustakas}, J. and {Mueller}, E. and {Mundet}, J. and {Mu{\~n}oz-Guti{\'e}rrez}, A. and {Myers}, A.~D. and {Nadathur}, S. and {Napolitano}, L. and {Neveux}, R. and {Newman}, J.~A. and {Nie}, J. and {Nikutta}, R. and {Niz}, G. and {Norberg}, P. and {Noriega}, H.~E. and {Paillas}, E. and {Palanque-Delabrouille}, N. and {Palmese}, A. and {Pan}, Z. and {Parkinson}, D. and {Penmetsa}, S. and {Percival}, W.~J. and {P{\'e}rez-Fern{\'a}ndez}, A. and {P{\'e}rez-R{\`a}fols}, I. and {Pieri}, M. and {Poppett}, C. and {Porredon}, A. and {Pothier}, S.},
        title = "{The Early Data Release of the Dark Energy Spectroscopic Instrument}",
      journal = {\aj},
         year = 2024,
        month = aug,
       volume = {168},
       number = {2},
          eid = {58},
        pages = {58},
          doi = {10.3847/1538-3881/ad3217},
archivePrefix = {arXiv},
       eprint = {2306.06308},
 primaryClass = {astro-ph.CO},
       adsurl = {https://ui.adsabs.harvard.edu/abs/2024AJ....168...58D}
}

@ARTICLE{fan22,
       author = {{Fan}, Xiaohui and {Ba{\~n}ados}, Eduardo and {Simcoe}, Robert A.},
        title = "{Quasars and the Intergalactic Medium at Cosmic Dawn}",
      journal = {\araa},
         year = 2023,
        month = aug,
       volume = {61},
        pages = {373-426},
          doi = {10.1146/annurev-astro-052920-102455},
archivePrefix = {arXiv},
       eprint = {2212.06907},
 primaryClass = {astro-ph.GA},
       adsurl = {https://ui.adsabs.harvard.edu/abs/2023ARA&A..61..373F}
}

@ARTICLE{boutsia20,
       author = {{Boutsia}, Konstantina and {Grazian}, Andrea and {Calderone}, Giorgio and {Cristiani}, Stefano and {Cupani}, Guido and {Guarneri}, Francesco and {Fontanot}, Fabio and {Amorin}, Ricardo and {D'Odorico}, Valentina and {Giallongo}, Emanuele and {Salvato}, Mara and {Omizzolo}, Alessandro and {Romano}, Michael and {Menci}, Nicola},
        title = "{The Spectroscopic Follow-up of the QUBRICS Bright Quasar Survey}",
      journal = {\apjs},
         year = 2020,
        month = oct,
       volume = {250},
       number = {2},
          eid = {26},
        pages = {26},
          doi = {10.3847/1538-4365/abafc1},
archivePrefix = {arXiv},
       eprint = {2008.03865},
 primaryClass = {astro-ph.GA},
       adsurl = {https://ui.adsabs.harvard.edu/abs/2020ApJS..250...26B}
}

@ARTICLE{grazian22,
       author = {{Grazian}, Andrea and {Giallongo}, Emanuele and {Boutsia}, Konstantina and {Calderone}, Giorgio and {Cristiani}, Stefano and {Cupani}, Guido and {Fontanot}, Fabio and {Guarneri}, Francesco and {Ozdalkiran}, Yacob},
        title = "{The Space Density of Ultra-luminous QSOs at the End of Reionization Epoch by the QUBRICS Survey and the AGN Contribution to the Hydrogen Ionizing Background}",
      journal = {\apj},
         year = 2022,
        month = jan,
       volume = {924},
       number = {2},
          eid = {62},
        pages = {62},
          doi = {10.3847/1538-4357/ac33a4},
archivePrefix = {arXiv},
       eprint = {2110.13736},
 primaryClass = {astro-ph.GA},
       adsurl = {https://ui.adsabs.harvard.edu/abs/2022ApJ...924...62G}
}

@ARTICLE{PSDR2,
       author = {{Flewelling}, H.~A. and {Magnier}, E.~A. and {Chambers}, K.~C. and {Heasley}, J.~N. and {Holmberg}, C. and {Huber}, M.~E. and {Sweeney}, W. and {Waters}, C.~Z. and {Calamida}, A. and {Casertano}, S. and {Chen}, X. and {Farrow}, D. and {Hasinger}, G. and {Henderson}, R. and {Long}, K.~S. and {Metcalfe}, N. and {Narayan}, G. and {Nieto-Santisteban}, M.~A. and {Norberg}, P. and {Rest}, A. and {Saglia}, R.~P. and {Szalay}, A. and {Thakar}, A.~R. and {Tonry}, J.~L. and {Valenti}, J. and {Werner}, S. and {White}, R. and {Denneau}, L. and {Draper}, P.~W. and {Hodapp}, K.~W. and {Jedicke}, R. and {Kaiser}, N. and {Kudritzki}, R.~P. and {Price}, P.~A. and {Wainscoat}, R.~J. and {Chastel}, S. and {McLean}, B. and {Postman}, M. and {Shiao}, B.},
        title = "{The Pan-STARRS1 Database and Data Products}",
      journal = {\apjs},
         year = 2020,
        month = nov,
       volume = {251},
       number = {1},
          eid = {7},
        pages = {7},
          doi = {10.3847/1538-4365/abb82d},
archivePrefix = {arXiv},
       eprint = {1612.05243},
 primaryClass = {astro-ph.IM},
       adsurl = {https://ui.adsabs.harvard.edu/abs/2020ApJS..251....7F}
}

@ARTICLE{PS1,
       author = {{Chambers}, K.~C. and {Magnier}, E.~A. and {Metcalfe}, N. and {Flewelling}, H.~A. and {Huber}, M.~E. and {Waters}, C.~Z. and {Denneau}, L. and {Draper}, P.~W. and {Farrow}, D. and {Finkbeiner}, D.~P. and {Holmberg}, C. and {Koppenhoefer}, J. and {Price}, P.~A. and {Rest}, A. and {Saglia}, R.~P. and {Schlafly}, E.~F. and {Smartt}, S.~J. and {Sweeney}, W. and {Wainscoat}, R.~J. and {Burgett}, W.~S. and {Chastel}, S. and {Grav}, T. and {Heasley}, J.~N. and {Hodapp}, K.~W. and {Jedicke}, R. and {Kaiser}, N. and {Kudritzki}, R. -P. and {Luppino}, G.~A. and {Lupton}, R.~H. and {Monet}, D.~G. and {Morgan}, J.~S. and {Onaka}, P.~M. and {Shiao}, B. and {Stubbs}, C.~W. and {Tonry}, J.~L. and {White}, R. and {Ba{\~n}ados}, E. and {Bell}, E.~F. and {Bender}, R. and {Bernard}, E.~J. and {Boegner}, M. and {Boffi}, F. and {Botticella}, M.~T. and {Calamida}, A. and {Casertano}, S. and {Chen}, W. -P. and {Chen}, X. and {Cole}, S. and {Deacon}, N. and {Frenk}, C. and {Fitzsimmons}, A. and {Gezari}, S. and {Gibbs}, V. and {Goessl}, C. and {Goggia}, T. and {Gourgue}, R. and {Goldman}, B. and {Grant}, P. and {Grebel}, E.~K. and {Hambly}, N.~C. and {Hasinger}, G. and {Heavens}, A.~F. and {Heckman}, T.~M. and {Henderson}, R. and {Henning}, T. and {Holman}, M. and {Hopp}, U. and {Ip}, W. -H. and {Isani}, S. and {Jackson}, M. and {Keyes}, C.~D. and {Koekemoer}, A.~M. and {Kotak}, R. and {Le}, D. and {Liska}, D. and {Long}, K.~S. and {Lucey}, J.~R. and {Liu}, M. and {Martin}, N.~F. and {Masci}, G. and {McLean}, B. and {Mindel}, E. and {Misra}, P. and {Morganson}, E. and {Murphy}, D.~N.~A. and {Obaika}, A. and {Narayan}, G. and {Nieto-Santisteban}, M.~A. and {Norberg}, P. and {Peacock}, J.~A. and {Pier}, E.~A. and {Postman}, M. and {Primak}, N. and {Rae}, C. and {Rai}, A. and {Riess}, A. and {Riffeser}, A. and {Rix}, H.~W. and {R{\"o}ser}, S. and {Russel}, R. and {Rutz}, L. and {Schilbach}, E. and {Schultz}, A.~S.~B. and {Scolnic}, D. and {Strolger}, L. and {Szalay}, A. and {Seitz}, S. and {Small}, E. and {Smith}, K.~W. and {Soderblom}, D.~R. and {Taylor}, P. and {Thomson}, R. and {Taylor}, A.~N. and {Thakar}, A.~R. and {Thiel}, J. and {Thilker}, D. and {Unger}, D. and {Urata}, Y. and {Valenti}, J. and {Wagner}, J. and {Walder}, T. and {Walter}, F. and {Watters}, S.~P. and {Werner}, S. and {Wood-Vasey}, W.~M. and {Wyse}, R.},
        title = "{The Pan-STARRS1 Surveys}",
      journal = {arXiv e-prints},
         year = 2016,
        month = dec,
          eid = {arXiv:1612.05560},
        pages = {arXiv:1612.05560},
archivePrefix = {arXiv},
       eprint = {1612.05560},
 primaryClass = {astro-ph.IM},
       adsurl = {https://ui.adsabs.harvard.edu/abs/2016arXiv161205560C}
}

@ARTICLE{calderone19,
       author = {{Calderone}, Giorgio and {Boutsia}, Konstantina and
         {Cristiani}, Stefano and {Grazian}, Andrea and {Amorin}, Ricardo and
         {D{\textquoteright}Odorico}, Valentina and {Cupani}, Guido and
         {Fontanot}, Fabio and {Salvato}, Mara},
        title = "{Finding the Brightest Cosmic Beacons in the Southern Hemisphere}",
      journal = {\apj},
         year = "2019",
        month = "Dec",
       volume = {887},
       number = {2},
          eid = {268},
        pages = {268},
          doi = {10.3847/1538-4357/ab510a},
archivePrefix = {arXiv},
       eprint = {1909.06391},
 primaryClass = {astro-ph.IM},
       adsurl = {https://ui.adsabs.harvard.edu/abs/2019ApJ...887..268C}
}

@ARTICLE{Sch19a,
       author = {{Schindler}, Jan-Torge and {Fan}, Xiaohui and {McGreer}, Ian D. and
         {Yang}, Jinyi and {Wang}, Feige and {Green}, Richard and
         {Fynbo}, Johan P.~U. and {Krogager}, Jens-Kristian and
         {Green}, Elisabeth M. and {Huang}, Yun-Hsin and {Kadowaki}, Jennifer and
         {Patej}, Anna and {Wu}, Ya-Lin and {Yue}, Minghao},
        title = "{The Extremely Luminous Quasar Survey in the Sloan Digital Sky Survey Footprint. III. The South Galactic Cap Sample and the Quasar Luminosity Function at Cosmic Noon}",
      journal = {\apj},
         year = "2019",
        month = "Feb",
       volume = {871},
       number = {2},
          eid = {258},
        pages = {258},
          doi = {10.3847/1538-4357/aaf86c},
archivePrefix = {arXiv},
       eprint = {1812.04639},
 primaryClass = {astro-ph.GA},
       adsurl = {https://ui.adsabs.harvard.edu/abs/2019ApJ...871..258S}
}

@ARTICLE{Sch19b,
       author = {{Schindler}, Jan-Torge and {Fan}, Xiaohui and {Huang}, Yun-Hsin and
         {Yue}, Minghao and {Yang}, Jinyi and {Hall}, Patrick B. and
         {Wenzl}, Lukas and {Hughes}, Allison and {Litke}, Katrina C. and
         {Rees}, Jon M.},
        title = "{The Extremely Luminous Quasar Survey in the Pan-STARRS 1 Footprint (PS-ELQS)}",
      journal = {\apjs},
         year = "2019",
        month = "Jul",
       volume = {243},
       number = {1},
          eid = {5},
        pages = {5},
          doi = {10.3847/1538-4365/ab20d0},
archivePrefix = {arXiv},
       eprint = {1905.04069},
 primaryClass = {astro-ph.GA},
       adsurl = {https://ui.adsabs.harvard.edu/abs/2019ApJS..243....5S}
}

@ARTICLE{Wri10,
       author = {{Wright}, Edward L. and {Eisenhardt}, Peter R.~M. and {Mainzer}, Amy K. and
         {Ressler}, Michael E. and {Cutri}, Roc M. and {Jarrett}, Thomas and
         {Kirkpatrick}, J. Davy and {Padgett}, Deborah and
         {McMillan}, Robert S. and {Skrutskie}, Michael and {Stanford}, S.~A. and
         {Cohen}, Martin and {Walker}, Russell G. and {Mather}, John C. and
         {Leisawitz}, David and {Gautier}, Thomas N., III and {McLean}, Ian and
         {Benford}, Dominic and {Lonsdale}, Carol J. and {Blain}, Andrew and
         {Mendez}, Bryan and {Irace}, William R. and {Duval}, Valerie and
         {Liu}, Fengchuan and {Royer}, Don and {Heinrichsen}, Ingolf and
         {Howard}, Joan and {Shannon}, Mark and {Kendall}, Martha and
         {Walsh}, Amy L. and {Larsen}, Mark and {Cardon}, Joel G. and
         {Schick}, Scott and {Schwalm}, Mark and {Abid}, Mohamed and
         {Fabinsky}, Beth and {Naes}, Larry and {Tsai}, Chao-Wei},
        title = "{The Wide-field Infrared Survey Explorer (WISE): Mission Description and Initial On-orbit Performance}",
      journal = {\aj},
         year = "2010",
        month = "Dec",
       volume = {140},
       number = {6},
        pages = {1868-1881},
          doi = {10.1088/0004-6256/140/6/1868},
archivePrefix = {arXiv},
       eprint = {1008.0031},
 primaryClass = {astro-ph.IM},
       adsurl = {https://ui.adsabs.harvard.edu/abs/2010AJ....140.1868W}
}

@ARTICLE{Guarneri21,
       author = {{Guarneri}, Francesco and {Calderone}, Giorgio and {Cristiani}, Stefano and {Fontanot}, Fabio and {Boutsia}, Konstantina and {Cupani}, Guido and {Grazian}, Andrea and {D'Odorico}, Valentina},
        title = "{The probabilistic random forest applied to the selection of quasar candidates in the QUBRICS survey}",
      journal = {\mnras},
         year = 2021,
        month = sep,
       volume = {506},
       number = {2},
        pages = {2471-2481},
          doi = {10.1093/mnras/stab1867},
archivePrefix = {arXiv},
       eprint = {2106.12990},
 primaryClass = {astro-ph.IM},
       adsurl = {https://ui.adsabs.harvard.edu/abs/2021MNRAS.506.2471G}
}

@ARTICLE{des21,
       author = {{Abbott}, T.~M.~C. and {Adam{\'o}w}, M. and {Aguena}, M. and {Allam}, S. and {Amon}, A. and {Annis}, J. and {Avila}, S. and {Bacon}, D. and {Banerji}, M. and {Bechtol}, K. and {Becker}, M.~R. and {Bernstein}, G.~M. and {Bertin}, E. and {Bhargava}, S. and {Bridle}, S.~L. and {Brooks}, D. and {Burke}, D.~L. and {Carnero Rosell}, A. and {Carrasco Kind}, M. and {Carretero}, J. and {Castander}, F.~J. and {Cawthon}, R. and {Chang}, C. and {Choi}, A. and {Conselice}, C. and {Costanzi}, M. and {Crocce}, M. and {da Costa}, L.~N. and {Davis}, T.~M. and {De Vicente}, J. and {DeRose}, J. and {Desai}, S. and {Diehl}, H.~T. and {Dietrich}, J.~P. and {Drlica-Wagner}, A. and {Eckert}, K. and {Elvin-Poole}, J. and {Everett}, S. and {Evrard}, A.~E. and {Ferrero}, I. and {Fert{\'e}}, A. and {Flaugher}, B. and {Fosalba}, P. and {Friedel}, D. and {Frieman}, J. and {Garc{\'\i}a-Bellido}, J. and {Gaztanaga}, E. and {Gelman}, L. and {Gerdes}, D.~W. and {Giannantonio}, T. and {Gill}, M.~S.~S. and {Gruen}, D. and {Gruendl}, R.~A. and {Gschwend}, J. and {Gutierrez}, G. and {Hartley}, W.~G. and {Hinton}, S.~R. and {Hollowood}, D.~L. and {Honscheid}, K. and {Huterer}, D. and {James}, D.~J. and {Jeltema}, T. and {Johnson}, M.~D. and {Kent}, S. and {Kron}, R. and {Kuehn}, K. and {Kuropatkin}, N. and {Lahav}, O. and {Li}, T.~S. and {Lidman}, C. and {Lin}, H. and {MacCrann}, N. and {Maia}, M.~A.~G. and {Manning}, T.~A. and {Maloney}, J.~D. and {March}, M. and {Marshall}, J.~L. and {Martini}, P. and {Melchior}, P. and {Menanteau}, F. and {Miquel}, R. and {Morgan}, R. and {Myles}, J. and {Neilsen}, E. and {Ogando}, R.~L.~C. and {Palmese}, A. and {Paz-Chinch{\'o}n}, F. and {Petravick}, D. and {Pieres}, A. and {Plazas}, A.~A. and {Pond}, C. and {Rodriguez-Monroy}, M. and {Romer}, A.~K. and {Roodman}, A. and {Rykoff}, E.~S. and {Sako}, M. and {Sanchez}, E. and {Santiago}, B. and {Scarpine}, V. and {Serrano}, S. and {Sevilla-Noarbe}, I. and {Smith}, J. Allyn and {Smith}, M. and {Soares-Santos}, M. and {Suchyta}, E. and {Swanson}, M.~E.~C. and {Tarle}, G. and {Thomas}, D. and {To}, C. and {Tremblay}, P.~E. and {Troxel}, M.~A. and {Tucker}, D.~L. and {Turner}, D.~J. and {Varga}, T.~N. and {Walker}, A.~R. and {Wechsler}, R.~H. and {Weller}, J. and {Wester}, W. and {Wilkinson}, R.~D. and {Yanny}, B. and {Zhang}, Y. and {Nikutta}, R. and {Fitzpatrick}, M. and {Jacques}, A. and {Scott}, A. and {Olsen}, K. and {Huang}, L. and {Herrera}, D. and {Juneau}, S. and {Nidever}, D. and {Weaver}, B.~A. and {Adean}, C. and {Correia}, V. and {de Freitas}, M. and {Freitas}, F.~N. and {Singulani}, C. and {Vila-Verde}, G. and {Linea Science Server}},
        title = "{The Dark Energy Survey Data Release 2}",
      journal = {\apjs},
         year = 2021,
        month = aug,
       volume = {255},
       number = {2},
          eid = {20},
        pages = {20},
          doi = {10.3847/1538-4365/ac00b3},
archivePrefix = {arXiv},
       eprint = {2101.05765},
 primaryClass = {astro-ph.IM},
       adsurl = {https://ui.adsabs.harvard.edu/abs/2021ApJS..255...20A}
}

@ARTICLE{Akiyama18,
       author = {{Akiyama}, Masayuki and {He}, Wanqiu and {Ikeda}, Hiroyuki and {Niida}, Mana and {Nagao}, Tohru and {Bosch}, James and {Coupon}, Jean and {Enoki}, Motohiro and {Imanishi}, Masatoshi and {Kashikawa}, Nobunari and {Kawaguchi}, Toshihiro and {Komiyama}, Yutaka and {Lee}, Chien-Hsiu and {Matsuoka}, Yoshiki and {Miyazaki}, Satoshi and {Nishizawa}, Atsushi J. and {Oguri}, Masamune and {Ono}, Yoshiaki and {Onoue}, Masafusa and {Ouchi}, Masami and {Schulze}, Andreas and {Silverman}, John D. and {Tanaka}, Manobu M. and {Tanaka}, Masayuki and {Terashima}, Yuichi and {Toba}, Yoshiki and {Ueda}, Yoshihiro},
        title = "{The quasar luminosity function at redshift 4 with the Hyper Suprime-Cam Wide Survey}",
      journal = {\pasj},
         year = 2018,
        month = jan,
       volume = {70},
          eid = {S34},
        pages = {S34},
          doi = {10.1093/pasj/psx091},
archivePrefix = {arXiv},
       eprint = {1704.05996},
 primaryClass = {astro-ph.GA},
       adsurl = {https://ui.adsabs.harvard.edu/abs/2018PASJ...70S..34A}
}

@ARTICLE{Giallongo15,
       author = {{Giallongo}, E. and {Grazian}, A. and {Fiore}, F. and {Fontana}, A. and {Pentericci}, L. and {Vanzella}, E. and {Dickinson}, M. and {Kocevski}, D. and {Castellano}, M. and {Cristiani}, S. and {Ferguson}, H. and {Finkelstein}, S. and {Grogin}, N. and {Hathi}, N. and {Koekemoer}, A.~M. and {Newman}, J.~A. and {Salvato}, M.},
        title = "{Faint AGNs at z > 4 in the CANDELS GOODS-S field: looking for contributors to the reionization of the Universe}",
      journal = {\aap},
         year = 2015,
        month = jun,
       volume = {578},
          eid = {A83},
        pages = {A83},
          doi = {10.1051/0004-6361/201425334},
archivePrefix = {arXiv},
       eprint = {1502.02562},
 primaryClass = {astro-ph.CO},
       adsurl = {https://ui.adsabs.harvard.edu/abs/2015A&A...578A..83G}
}

@BOOK{CCA,
       author = {{Anderson}, T.W.},
        title = "{An introduction to multivariate statistical analysis}",
         year = 1984,
    publisher = {Wiley Ed.}
}
\bibliographystyle{aasjournal}

\begin{appendix}

\section{Selection of the band}

\begin{figure}
\includegraphics[width=\linewidth]{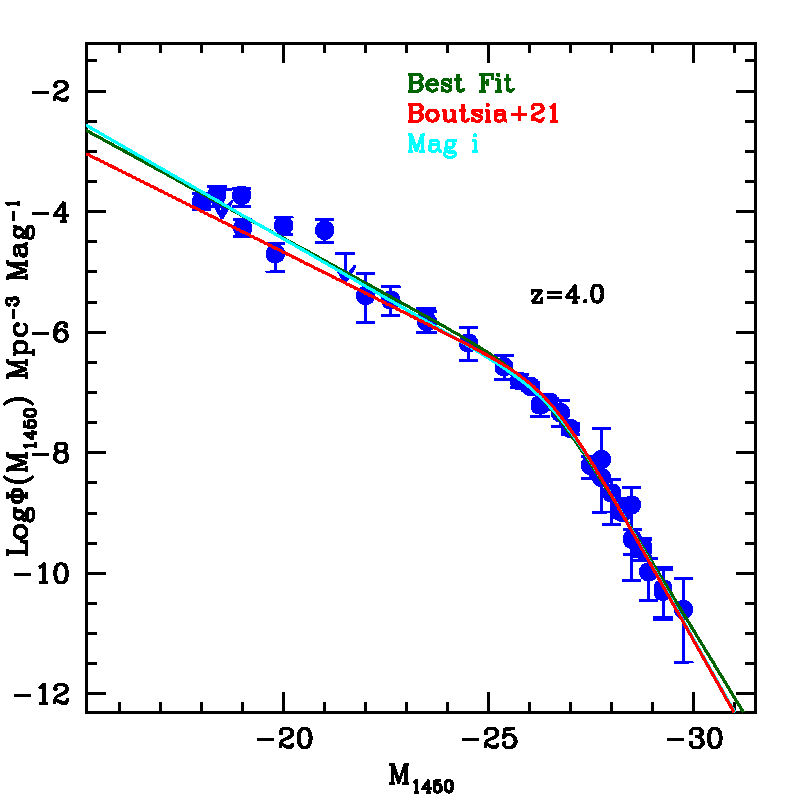}
\caption{The comparison of the $z\sim 4$ AGN LF from this work, based on the Y-band magnitude (green) and the one calculated, considering the i-magnitude for all sources (cyan). Practically, the two fits are identical and at the faint end they are both slightly steeper than the slope determined by \cite{boutsia21}. The blue points are the compilation of all surveys as seen in Fig\ref{fig:lffaint} and have been calculated based on the i-band.}
\label{fig:lfi}
\end{figure}

Since the goal of the QUBRICS survey is currently the search for high redshift QSOs ($z>5$), the Main Sample and the various selections are being created based on the Y-band. Thus this is the band in which the Sample is better constrained. Moreover, since the adopted completeness correction is only a constant factor at all magnitudes, the choice of the Y-band filter is not affecting the shape of the QSO LF in this work.
To check this assumption, we calculated the LF using the i-band magnitudes retrieved for all the sources. 
The best fit parameters are presented in Table \ref{tab:lffiti} and both sets of values agree within the $1-\sigma$ uncertainty. As can be seen also in Fig.\ref{fig:lfi}, the two LFs are basically identical.

\begin{table}
\caption{The best fit parameters of the QSO luminosity function at
$3.6\le z\le 4.2$.}
\label{tab:lffiti}
\begin{center}
\begin{tabular}{c c c c c c c}
\hline
\hline
Band&$\alpha$ & $\beta$ & $M^*_{1450}$ & $Log\Phi^*$ \\
\hline
i &$-1.98_{-0.09}^{+0.07}$ & $-4.02_{-0.38}^{+0.43}$ & $-26.66_{-0.41}^{+0.62}$
& $ -7.03_{-0.32}^{+0.45}$ \\
Y & $-1.93_{-0.08}^{+0.10}$ & $-3.80_{-0.37}^{+0.28}$ & $-26.27_{-0.51}^{+0.63}$
& $ -6.74_{-0.39}^{+0.46}$ \\
\hline
\hline
\end{tabular}
\end{center}
The errors associated to the best fit parameters are at 68\% confidence level (1 $\sigma$).
\end{table}

\begin{table*}
\caption{The confirmed QSOs at $z\sim 4$ from DES and PanSTARRS in the total area (global) used for the final LF fit ($M_{1450} \leq$-27.0).}
\label{tab:glob}
\begin{center}
\begin{tabular}{l c c c c c c}
\hline
QUBRICS ID & RA & DEC & $z_{spec}$ & i-band mag. & $Y-band$ mag. & $M_{1450}$  \\
   1122453  &      38.989861031  &      -34.81570692 &    3.74 & 17.69  &  17.39 &  -28.26 \\
  1122900  &      31.055212747  &      -32.85634969 &    3.83 & 17.03  &  16.94 &  -28.77 \\
  36004563 &      35.497708387  &      -32.69828351 &    3.68 & 18.28  &  17.59 &  -28.03 \\
  1077319  &      61.938703462  &      -32.26047362 &    3.75 & 17.67  &  17.19 &  -28.47 \\
... & ... & ... & ... &  ... &... & ... \\
  1144132  &      89.00269554   &      -32.7202966  &    4.04 & 18.02  &  17.88 &     -27.95 \\
  1034896  &      21.546804042  &      -31.81164919 &    4.03 & 17.91  &  17.95 &     -27.87 \\
  1000231  &      21.820331809  &      -58.0464918  &    3.92 & 17.81  &  17.79 &     -27.98 \\
  40867501 &      327.30971973  &      -39.75797936 &    3.97 & 18.75  &  18.56 &     -27.23 \\
\hline
\hline
\end{tabular}
\tablecomments{
Full table to be found on the electronic version of the paper. 
\\
}
\end{center}
\end{table*}

\begin{table*}
\caption{The confirmed QSOs at $z\sim 4$ in the DESI area used for the LF.}
\label{tab:desi}
\begin{center}
\begin{tabular}{l c c c l c c c}
\hline
DESI ID & RA & DEC & z$_{spec}$ & PanSTARRS ID & $i-band$ mag. & $Y-band$ mag.& $M_{1450}$  \\
39627770183026111(1)  &  213.06042790 & -0.79243674  &   3.76 &  107042130604669541   & 19.93  &  19.74  &  -25.93 \\
39627788306613340(2)  &  213.31402466 &  0.00899046  &   4.08 &  108012133139031167   & 19.53  &  19.51  &  -26.35 \\
39627927020637069(3)  &  210.26305889 &  5.65395372  &   3.71 &  114782102630645215   & 19.94  &  19.56  &  -26.08 \\
39627734002963089     &  216.70572360 & -2.13932657  &   3.69 &  105432167057613211   & 20.18  &  19.85  &  -25.77 \\ 
  ... & ... & ... & ... & ... & ... & ... & ... \\
39627951066579939  &  210.72310204 &  6.67091859  &   3.88 &  116002107231265522      & 20.14  &  20.05  &  -25.69  \\
39627764168393687  &  214.54027705 & -1.08398580  &   3.98 &  106692145402489687      & 20.36  &  19.79  &  -26.01  \\
No DESI ID         &  215.52343978 & -1.47011544  &   3.91 &  106232155235506358      & 20.74  &  20.11  &  -25.65  \\
No DESI ID         &  218.9807624  &  1.93069167  &   3.60 &  110312189807587194      & 20.48  &  20.21  &  -25.37  \\
\hline
\hline
\end{tabular}
\tablecomments{
\\
The spectroscopically confirmed QSOs at $3.6<z<4.2$ in the five selected pointings of the DESI survey, as described in the main text. The first 3 sources have two distinct IDs in the DESI database. \\
Full table to be found on the electronic version of the paper. . 
\\
}
\end{center}
\end{table*}

\end{appendix}

\end{document}